\documentclass{jfm}
\usepackage{graphicx}
\usepackage{epstopdf, epsfig}

\usepackage[normalem]{ulem}

\usepackage{orcidlink}

\usepackage{amsmath,amssymb, bm}
\usepackage{textcomp}
\usepackage{amssymb}
\newcommand{\polarorder}{\bm{m}}
\DeclareMathOperator{\sech}{sech}
\renewcommand{\Re}{\operatorname{Re}}

\newcommand{\bOmega}{\bm{\Omega}}

\newcommand{\be}{\bm{e}}

\newcommand{\bx}{\bm{x}}

\newcommand{\bu}{\bm{u}}

\newcommand{\bq}{\bm{q}}

\newcommand{\odiff}{\text{d}} 
\newcommand{\intS}{\int_{\mathbb{S}}}

\newcommand{\dx}{\odiff x}
\newcommand{\dy}{\odiff y}

\newcommand{\dq}{\odiff\bq}
\newcommand{\dphi}{\odiff\phi}
\newcommand{\avg}[1]{\left< #1 \right>}

\newcommand{\bj}{\bm{j}}
\usepackage{xcolor}

{} 

\shorttitle{Active particles in oscillatory flow}
\shortauthor{V. Chakraborty, P. Mishra, M. Qiu and Z. Peng} 

\title{Dispersion of active particles in oscillatory Poiseuille flow}

\author{
    Vhaskar Chakraborty\aff{1}, Pankaj Mishra\aff{1}, Mingfeng Qiu\aff{2}, \and Zhiwei Peng\,\orcidlink{0000-0002-9486-2837 }\aff{3}\corresp{\email{zhiwei.peng@ualberta.ca}}
}

\affiliation{
    \aff{1} Department of Physics, Indian Institute of Technology (ISM), Dhanbad  
826004, India\\
    \aff{2} School of Mathematics and Statistics, University of Canterbury, Christchurch 8140, New Zealand\\
    \aff{3} Department of Chemical and Materials Engineering, University of Alberta, Edmonton, Alberta T6G 1H9, Canada
}

\begin{document}

\maketitle

\begin{abstract}
Active particles exhibit complex transport dynamics in flows through confined geometries such as channels or pores. In this work, we employ a generalized Taylor dispersion (GTD) theory to study the long-time dispersion behavior of active Brownian particles (ABPs) in an oscillatory Poiseuille flow within a planar channel. We quantify the time-averaged longitudinal dispersion coefficient as a function of the flow speed, flow oscillation frequency, and particle activity. In the weak-activity limit, asymptotic analysis shows that activity can either enhance or hinder the dispersion compared to the passive case. For arbitrary activity levels, we numerically solve the GTD equations and validate the results with Brownian dynamics simulations. We show that the dispersion coefficient could vary non-monotonically with both the flow speed and particle activity. Furthermore, the dispersion coefficient shows an oscillatory behavior as a function of the flow oscillation frequency, exhibiting distinct minima and maxima at different frequencies. The observed oscillatory dispersion results from the interplay between self-propulsion and oscillatory flow advection---a coupling absent in passive or steady systems. Our results show that time-dependent flows can be used to tune the dispersion of active particles in confinement.

\end{abstract}


\section{\label{sec :intro}Introduction}

For micron-sized particles, the presence of fluid flow can enhance mass transport due to the interplay between advection and diffusion. A classical example of this coupling effect is Taylor dispersion, where Brownian solutes in pressure-driven flows exhibit enhanced longitudinal dispersion compared to the molecular diffusivity \citep{Taylor3, Taylor, Taylor2,aris1956dispersion}. 
Since the work of \citet{Taylor3}, a generalized Taylor dispersion (GTD) framework has been developed to study a variety of transport phenomena. These include complex geometries, spatial and temporal periodicity, and active (i.e., self-propelled) particle dynamics \citep{brenner1980dispersion,shapiro1990taylor,hill2002taylor,zia2010single,brenner2013macrotransport,alonso2019transport,peng2020upstream,peng2024macrotransport}.

Active particles differ from passive solutes in that each unit is capable of self-propulsion \citep{schweitzer1998complex,romanczuk2012active}.
The interplay between self-propulsion and fluid flow gives rise to rich and often non-intuitive dynamics that are absent in passive Brownian systems \citep{romanczuk2012active, bechinger2016active, gomez2016,plan2020active,jing2020chirality,Chandragiri2020, chakraborty2022active,  choudhary2022inertial}. One example where these dynamics play a crucial role is the transport behavior of microswimmers, which is important for understanding both natural and engineered systems, such as infection by motile bacteria \citep{Siitonen1992,lane2005role}, formation of biofilms \citep{kim2014filaments,rusconi2010laminar}, drug delivery \citep{Park2017, Lin2020, Diez2021,Sridhar2022}, therapeutic treatments \citep{ Ghosh2020} and environmental remediation \citep{Soler2013,Urso2023}. 

Transport of active particles often occur in confined geometries, where Poiseuille flow is a common flow profile, and considerable work has focused on how active matter behaves in such environments 
\citep{zottl2012nonlinear, zottl2013periodic,apaza2016ballistic, junot2019swimming, mathijssen2019oscillatory, chuphal2021effect,anand2021migration,
khatri2022diffusion, choudhary2022inertial, walker2022emergent, ganesh2023numerical,valani2024active}. For instance, in channels, active particles exhibit upstream swimming in Poiseuille flow
\citep{kaya2012direct,kantsler2014rheotaxis,ezhilan2015transport,Omori2016}.
Owing to their upstream motility, \textit{E. ~coli} introduced downstream causes upstream contamination in initially clean microfluidic channels \citep{figueroa2020coli}. Further investigations by \citet{mathijssen2019oscillatory} on bacterial motion near channel surfaces revealed that \textit{E.~coli} engages in distinct rheotaxis regimes depending on the shear rate. With increasing shear, the bacteria transition from upstream swimming to oscillatory rheotaxis, and ultimately to a coexistence of rheotaxis aligned with both positive and negative vorticity directions.

While these studies were primarily focused on steady flows, biologically relevant systems are often governed by time-dependent flow conditions. \citet{McDonald1955} experimentally studied the relationship between pulsatile pressure and blood flow in arteries, analyzing the phasic variations in arterial flow during each cardiac cycle. Inspired by the study of \citet{McDonald1955}, \citet{womersley1955method} investigated the velocity, rate of flow, and viscous drag in arteries by considering a time-periodic pressure gradient. The primary factors governing such flows include the pulsatile pressure generated by the heart, the structural and mechanical properties of the vascular walls, and the flow behavior of blood \citep{secomb2016hemodynamics}. 

Early studies on longitudinal dispersion of passive contaminants in oscillatory pressure-driven flows were carried out by \citet{Chatwin_1975,Chatwin_1977}. Later, \cite{Watson_1983} derived analytical solutions for the long-time effective dispersivity in oscillatory flows within both pipes and rectangular channels. His results showed that the effective dispersivity decreases monotonically with increasing flow frequency.
Subsequently, \citet{mazumder1992effect} investigated how boundary absorption and heterogeneous reactions influence contaminant dispersion in both steady and oscillatory flows. The significance of such boundary interactions lies in their relevance to processes such as deposition and transport across semi-permeable membranes.
More recently, \citet{chu2019dispersion} developed a macro-transport theory for two-dimensional flows in a parallel plate channel with alternating shear-free and no-slip regions.
They considered both steady and oscillatory flow components to study the transport coefficients of passive particles. Later, they extended their analysis to eccentric annuli \citep{chu2020dispersion} where they showed that the maximum dispersion observed in a time-oscillatory flow can be achieved by applying a slowly oscillating flow in an annulus with large eccentricity. 
\citet{hettiarachchi2011effect} used experiments and simulations to show that pulsatile cerebrospinal fluid significantly enhances drug dispersion in the spinal cord relative to no flow.

Although the dispersion of passive particles in oscillatory flows has been widely studied, much less is known about the transport of microswimmers in oscillatory flows. Recently, using experiments and simulations, \citet{caldag2025fine} showed that oscillatory flow can lead to nontrivial dispersion dynamics in gyrotactic swimmers. \citet{wang2025taylor} studied Taylor–Aris dispersion of active particles in oscillatory channel flows and showed that spherical non-gyrotactic swimmers can exhibit either enhanced or reduced diffusivity relative to passive solutes due to disruption of cross-streamline migration associated with Jeffery orbits. \citet{lagoin2025enhanced} experimentally investigated the motility and dispersion of \textit{Chlamydomonas reinhardtii} microalgae within a rectangular microfluidic channel under sinusoidal Poiseuille flow, showing that velocity fluctuations and the dispersion coefficient increase with flow amplitude, with weak dependencies on flow periodicity. In this paper, we consider the dispersion of active Brownian particles (ABPs) in time-periodic pressure-driven Poiseuille flow through planar channels. We apply the GTD theory of \citet{peng2020upstream}, originally developed for ABPs in steady flow, to characterize the long-time longitudinal dispersion of ABPs in oscillatory flow. Due to the time-periodic nature of the flow, an additional time average over one oscillation period is performed to define the time-averaged dispersion coefficient \citep{Chatwin_1975,Chatwin_1977,Watson_1983}. In the weak-swimming limit, characterized by a small swim P\'eclet number ($Pe_s \ll 1$), we show that the first effect of swimming on longitudinal dispersion appears at $O(Pe_s^2)$. Depending on the flow P\'eclet number ($Pe$) and oscillation frequency, the $O(Pe_s^2)$ contribution can be either positive or negative. As such, activity can either enhance or hinder longitudinal dispersion in oscillatory Poiseuille flow compared to passive Brownian particles.  For arbitrary swim speeds, numerical solutions of the governing equations are used to characterize the dispersion as a function of the flow speed, swim speed, and oscillation frequency.  Numerical results are validated against  Brownian dynamics (BD) simulations.

\section{Problem formulation}
\label{Sec :Probform}
\subsection{The Smoluchowski equation}
\label{Subsec :Smoluchowski}
We consider the long-time transport behavior of ABPs dispersed in a viscous Newtonian solvent confined between two parallel plates with a separation distance of $2H$. In the dilute limit, we only consider the dynamics of a single ABP. The ABP is assumed to be spherical, and its radius is much smaller than the width of the channel. This allows us to treat the ABP as a `point' particle. An ABP self-propels with a constant swim speed $U_s$ in a body-fixed swimming direction $\bq$ ($\bq\cdot\bq=1$). Due to rotational Brownian motion, the orientation vector $\bq$ undergoes stochastic reorientation. The configuration of an ABP at time $t$ is described by its position vector $\bx$ and by the orientation vector $\bq$. We define  $P(\bx, \bq, t)$ as the probability density function of finding the ABP at position $\bx$ with orientation $\bq$ at time $t$. It satisfies the Smoluchowski equation,
\begin{equation}
\label{eq:smol}
    \frac{\partial P}{\partial t} + \nabla \cdot\bj_T + \nabla_R \cdot \bj_R=0, 
\end{equation}
where $\nabla = \partial/\partial \bx$ and $\nabla_R = \bq \times \partial /\partial \bq$ are the spatial and rotational gradient operators, respectively. In equation \eqref{eq:smol}, 
\begin{equation}
    \bj_T = U_s \bq P + \bu_f P - D_T \nabla P, 
\end{equation}
\begin{equation}
    \bj_R = \bOmega_f P -D_R \nabla_R P, 
\end{equation}
where $\bu_f$ is the background fluid velocity field, $D_T$ is the translational diffusivity of the ABP, $\bOmega_f = \frac{1}{2}\nabla \times \bu_f$ is the flow-induced angular velocity, 
    and $D_R$ is the rotational diffusivity of the ABP. The inverse of $D_R$, $\tau_R = 1/D_R$, defines the reorientation time. At the channel walls, the no-flux boundary condition is satisfied \citep{ezhilan2015transport,peng2020upstream}:
\begin{equation}
\label{eq:bc-noflux}
   \bm{e}_y \cdot \bj_T =0, \quad y = \pm H, 
\end{equation}
where $\bm{e}_y$ is the unit normal to the channel walls. The longitudinal Cartesian coordinate is $x$ and  $y$ is the transverse coordinate.

\subsection{Oscillatory Poiseuille flow}
\label{Subsec :Oscillatory}
For ease of reference, we provide a brief outline of the flow field derivation. We consider a one-dimensional flow, $\bu_f = u(y,t)\be_x$, driven by a prescribed oscillatory pressure gradient along the channel \citep{womersley1955method}. Here $\be_x$ is the unit basis vector in the longitudinal direction.  The Navier-Stokes equations reduce to 
\begin{equation}
\label{eq:NS}
   \rho  \frac{\partial u}{\partial t}  = - \frac{\partial p}{\partial x} + \mu \frac{\partial^2 u }{\partial y^2}, 
\end{equation}
where $\rho$ is the density of the fluid, $\mu$ is the dynamic viscosity of the fluid, and the prescribed pressure gradient is given by 
\begin{equation}
\label{eq:dp-dx}
    - \frac{\partial p}{\partial x} = \frac{P_0}{H} \cos(\omega t).
\end{equation}
In equation~\eqref{eq:dp-dx}, $P_0$ is a reference pressure and $\omega$ is the angular frequency of the actuation. One can show that the solution of equation \eqref{eq:NS} may be written as $u(y,t) =\Re\left[ u^\prime(y) e^{i\omega t} \right] $, where 
\begin{equation}
\label{eq:u-prime}
    u^\prime(y) = \frac{i P_0}{\rho H \omega} \left[-1 + \cosh\left( (1+i)\lambda y \right) \sech\left((1+i)\lambda H \right)\right].
\end{equation}
In equation \eqref{eq:u-prime}, $i=\sqrt{-1}$ is the imaginary unit, $\nu = \mu/\rho$ is the kinematic viscosity of the fluid, and $\lambda = \sqrt{\omega/(2\nu)}$. The viscous length, $1/\lambda = \sqrt{2\nu/\omega}$, sets the scale over which the fluid momentum diffuses during one oscillation cycle of the applied pressure. The operator $\Re$ extracts the real part of a complex quantity. 

In the zero-frequency limit, $\omega \to 0$, we recover the steady Poiseuille flow as 
\begin{equation}
    u(y,t) \to \frac{P_0H}{2 \mu} \left(1 - \frac{y^2}{H^2} \right).
\end{equation}
For convenience, we define the characteristic flow speed $U_f = P_0H/(2\mu)$. Using this, we rewrite equation~\eqref{eq:u-prime} as 
\begin{equation}
      u^\prime(y) = \frac{i U_f}{(\lambda H)^2} \left[-1 + \cosh\left( (1+i)\lambda y \right) \sech\left((1+i)\lambda H \right)\right],  
\end{equation}
The angular velocity $\Omega_f(y, t) =  \Re\left[\Omega^\prime e^{i\omega t}\right] $, where 
\begin{equation}
    \Omega^\prime = - \frac{1}{2} \frac{\partial u^\prime}{\partial y} = \frac{(1-i) U_f}{2 \lambda H^2}  \sinh\left( (1+i)\lambda y \right) \sech\left((1+i)\lambda H \right). 
\end{equation}

\subsection{Generalized Taylor dispersion theory}
\label{Subsec :Taylor}
Taking the zeroth orientational moment of equation~\eqref{eq:smol} gives the governing equation for the number density, 
\begin{equation}
    \frac{\partial n}{\partial t} + \nabla \cdot\left( \bu_f n + U_s \polarorder - D_T \nabla n \right)=0,
\end{equation}
where  $n = \intS P \dq$ is the number density, and $\polarorder = \intS \bq P \dq $ is the first moment, or polar order. Here $\mathbb{S} = \{\bq \, | \, \bq\cdot\bq = 1\}$ denotes the unit sphere of orientations.
    In the following, we present a general derivation and then specialize to two dimensions (2D).  Since the channel is unbounded in the $x$ direction, it is convenient to work in Fourier space. To derive a long-time effective transport equation, we first define the Fourier transform of a function $f(x)$ as $\hat{f}(k)=\int e^{-ikx} f(x) \dx$, where $k$ is the wavenumber. Following \citet{peng2020upstream}, one can show that 
\begin{equation}
\label{eq:n-bar}
    \frac{\partial \overline{n}}{\partial t} + k^2 D_T \overline{n} + ik \left(\overline{u(y,t)\hat{n}} +U_s \overline{\hat{m}_x} \right)=0, 
\end{equation}
where we have made use of the no-flux condition, and an overhead bar denotes the cross-sectional average, 
\begin{equation}
    \overline{n}(k, t) = \frac{1}{2H}\int_{-H}^H \hat{n}(k,y,t) \dy.  
\end{equation}
In equation~\eqref{eq:n-bar}, $\hat{m}_x = \be_x\cdot \hat{\polarorder}$ is the polar order in the $x$ direction in Fourier space. 

Introducing the non-dimensional density or structure function $\hat{G}$ such that $\hat{P}(k, y, \bq, t) = \overline{n}(k,t)\hat{G}(k,y,\bq,t)$ and the small wavenumber expansion $\hat{G} = g(y, \bq, t) + ik\,  b(y, \bq, t) +O(k^2)$, we obtain 
\begin{equation}
\label{eq:small-k-adv-diffu}
    \frac{\partial \overline{n}}{\partial t}+ ik U^\mathrm{eff}\overline{n} + k^2 D^\mathrm{eff}\overline{n} + O(k^3)=0, 
\end{equation}
where the effective drift and the effective longitudinal dispersivity are given by, respectively, 
\begin{equation}
\label{eq:Ueff}
    U^\mathrm{eff} = U^\mathrm{eff}(t) = U_s \overline{m_x^0} + \overline{un^0}, 
\end{equation}
\begin{equation}
\label{eq:Deff}
    D^\mathrm{eff} = D^\mathrm{eff}(t) = D_T - U_s\overline{\tilde{m}_x}-\overline{u\tilde{n}}. 
\end{equation}
In the small-wavenumber expansion, $g$ is the average field and $b$ is the displacement (or fluctuating) field. Note that $b$ has units of length, e.g., displacement. We emphasize that terms of order $k^3$ and higher do not contribute to either the drift or the dispersion coefficient, as evidenced in equation \eqref{eq:small-k-adv-diffu}. In general, a diffusive flux is proportional to the gradient of a concentration; equivalently, in a small-wavenumber expansion in Fourier space, it appears at second order in $k$. Physically, a diffusivity describes only leading-order gradient transport and therefore cannot capture higher-order effects. Likewise, drift and diffusivity characterize only the lowest moments of the underlying probability distribution. To resolve the full probability distribution,  one must retain higher-order moments. The orientational moments in \eqref{eq:Ueff} are given by 
\begin{equation}
    n^0 = \intS g \dq, \quad \mathrm{and}\quad \polarorder^0=\intS \bq g\dq. 
\end{equation}
Similarly, in \eqref{eq:Deff}, we have 
\begin{equation}
    \tilde{n} = \intS b \dq, \quad \mathrm{and}\quad \tilde{\polarorder}=\intS \bq b\dq. 
\end{equation}
Different from the constant transport coefficients in \citet{peng2020upstream},  the long-time transport coefficients in \eqref{eq:Ueff} and \eqref{eq:Deff} are time-dependent due to the oscillatory flow.  

The governing equations and boundary conditions for $g$ and $b$ are derived in \citet{peng2020upstream}. For the average field, we have 
\begin{equation}
\label{eq:g-eq}
    \frac{\partial g}{\partial t} + \frac{\partial }{\partial y}\left(  U_s q_y g -D_T \frac{\partial g}{\partial y}\right) +\nabla_R\cdot\left( \bOmega_f g - D_R \nabla_R g\right)=0,
\end{equation}
and 
\begin{equation}
\label{eq:BC-1}
     U_s q_y g -D_T \frac{\partial g}{\partial y}=0, \quad y=\pm H. 
\end{equation}
The displacement field is governed by 
\begin{equation}
\label{eq:b-eq}
    \frac{\partial b}{\partial t} + \frac{\partial }{\partial y}\left(  U_s q_y b -D_T \frac{\partial b}{\partial y}\right) +\nabla_R\cdot\left( \bOmega_f b - D_R \nabla_R b\right)=\left(U^\mathrm{eff} - u-U_s q_x\right) g,
\end{equation}
\begin{equation}
\label{eq:BC-2}
 U_s q_y b -D_T \frac{\partial b}{\partial y}=0, \quad y=\pm H. 
\end{equation}
Noting that 
\begin{equation}
    \frac{1}{2H}\int_{-H}^H \dy \intS \hat{G}\dq=1, 
\end{equation}
we have 
\begin{equation}
\label{eq:g-b-integral}
        \frac{1}{2H}\int_{-H}^H \dy \intS g\dq=1, \quad\mathrm{and}\quad \frac{1}{2H}\int_{-H}^H \dy \intS b \dq=0. 
\end{equation}

    The above derivation of the GTD theory applies in both two and three dimensions. In the remainder of the paper, we restrict attention to two dimensions, where the orientation vector is parametrized as $\bq = \cos\phi\,\be_x + \sin\phi \,\be_y$,  with $\phi \in [0, 2\pi)$ being the orientation angle.  In 2D, the rotational gradient operator is given by $\nabla_R = \be_{z} \frac{\partial }{\partial \phi}$, where $\be_z = \be_x \times \be_y$.

\subsection{Non-dimensionalization}
\label{Subsec :Non-dimensionalization}
We scale lengths with the channel half-width $H$ and time with the reorientation time $\tau_R$.  The system is governed by five non-dimensional parameters:
\begin{subequations}
\begin{equation}
\label{eq:Dim-1}
    Pe = \frac{U_f \tau_R}{H}, \quad Pe_s = \frac{U_s \tau_R}{H} = \frac{\ell}{H}, \quad \gamma =  \frac{\sqrt{D_T\tau_R}}{H}=\frac{\delta}{H}, 
\end{equation}
\begin{equation}
\label{eq:Dim-2}
    \quad \chi = \omega\, \tau_R, \quad \kappa =\lambda H =  \sqrt{\omega/(2\nu)}H.
\end{equation}
\end{subequations}
where $Pe$ is the flow P\'eclet number that compares the reorientation time $\tau_R$ with the flow timescale $H/U_f$, $Pe_s$ is the swim P\'eclet number that compares the reorientation time with the swim timescale $H/U_s$, $\gamma$ is a non-dimensional measure of the microscopic length $\delta=\sqrt{D_T\tau_R}$, $\chi$ is the non-dimensional flow frequency, and $\kappa$ compares the length scale $1/\lambda$ with the channel half-width $H$. The microscopic length $\delta$ characterizes the distance a particle travels by translational diffusion over the timescale defined by $\tau_R$.  The swim P\'eclet number can be viewed as a comparison between the persistence length (or run length), $\ell=U_s\tau_R$, and the channel half-width.

Since both $\chi$ and $\kappa$ contains $\omega$, it is useful to introduce the non-dimensional parameter 
\begin{equation}
\label{eq: alpha}
    \alpha = \frac{\chi }{\kappa^2} = \frac{2\nu\tau_R}{H^{2}},
\end{equation}
when analyzing the effect of flow frequency $\omega$ on dispersion behavior. With this, varying the dimensional frequency $\omega$ corresponds to changing $\chi$ while keeping $\alpha$ constant. To estimate the order of magnitude of $\alpha$ in realistic systems, consider motile bacteria such as \textit{E. coli} in water at room temperature. Taking  $H \sim 100 ~\mu \text{m}$,  $\tau_R \sim 1~\text{s}$~\citep{berg1972chemotaxis,berg2004coli}, and $\nu \sim 10^{-6}~\text{m}^2/\text{s}$, we have $\alpha \sim 10^2$. For narrower microfluidic channels, $\alpha$ would be even larger. For bacteria or synthetic active particles with a shorter reorientation time, $\alpha$ is smaller. In the remainder of the paper, we take $\alpha = 100$ unless stated otherwise. 

The non-dimensional form of equation \eqref{eq:g-eq} is
\begin{equation}
\label{eq:g-non-dimensional}
    \frac{\partial g}{\partial t^*} + \frac{\partial }{\partial y^*}\left(  Pe_s q_y g - \gamma^2 \frac{\partial g}{\partial y^*}\right) +\frac{\partial}{\partial \phi}\left( \Omega_f^* g -  \frac{\partial g}{\partial \phi} \right)=0,
\end{equation}
where we have used the parametrization $\bq = \cos\phi\,\be_x + \sin\phi \,\be_y$ with $\phi \in [0, 2\pi)$ being the orientation angle, $y^*\in [-1,1]$, and we have used the superscript `$*$' to denote dimensionless quantities. That is, $t^* = t/\tau_R$, $y^*=y/H$, and $\Omega_f^*=\Omega_f \tau_R = \Re\left[ \Omega^{\prime *} e^{i \chi t^*} \right]$, where 
\begin{equation}
\label{eq: nondimangvel}
    \Omega^{\prime *} = \Omega^\prime \tau_R =\frac{(1-i) Pe}{2\kappa}\sinh\left( (1+i)\kappa y^* \right) \sech\left((1+i)\kappa \right).
\end{equation}
The superscript on $g$ is suppressed since $g$ is non-dimensional. With the solution of $g$, we can obtain the non-dimensional drift via 
\begin{equation}
\label{eq:Ueff-non-dimensional}
    U^{\mathrm{eff}*}(t^*) = U^\mathrm{eff}\tau_R/H = Pe_s \overline{m_x^0} + \overline{u^*n^0}. 
\end{equation}

Similarly, we may write the non-dimensional form of equation~\eqref{eq:b-eq} as 
\begin{equation}
\label{eq:b-non-dimensional}
        \frac{\partial b^*}{\partial t^*} + \frac{\partial }{\partial y^*}\left(  Pe_s q_y b^* -\gamma^2 \frac{\partial b^*}{\partial y^*}\right) +\frac{\partial }{\partial\phi}\left( \Omega^*_f b^* - \frac{\partial b^*}{\partial \phi}\right)=\left(U^{\mathrm{eff}*} - u^*-Pe_s q_x\right) g,
\end{equation}
where $b^* = b/H$, and $u^* =u \tau_R/H = \Re[u^{\prime *} e^{i \chi t^*} ]$. The complex flow amplitude is given by 
\begin{equation}
\label{eq: nondimlinvel}
    u^{\prime *} = \frac{i Pe}{\kappa^2 } \left[-1 + \cosh\left( (1+i)\kappa y^* \right) \sech\left((1+i)\kappa\right)\right]. 
\end{equation}

To characterize the dispersion of active particles in an oscillatory Poiseuille flow, we compare the effective dispersion coefficient with the translational diffusivity. Using equation \eqref{eq:Deff}, we have 
\begin{equation}
\label{eq:Deff_over_DT}
    D^\mathrm{eff*} = \frac{D^\mathrm{eff}}{D_T}= 1 - \frac{Pe_s}{\gamma^2}\overline{\tilde{m}_x^*} - \frac{1}{\gamma^2}\overline{u^*\tilde{n}^*}. 
\end{equation}
If $Pe=0$, or $U_f=0$, the problem reduces to that of diffusion of ABPs in a flat channel without flow. In this case, we have $D^\mathrm{eff}= D^\mathrm{eff}_\mathrm{nf}=D_T +D^\mathrm{swim}$, where $D^\mathrm{swim} = U_s^2\tau_R/2$ in 2D \citep{berg1993random}, and $ D^\mathrm{eff}_\mathrm{nf}$ is the effective dispersivity without flow. In non-dimensional form, we have
\begin{equation}
\label{eq: noflowdispersivity}
    \frac{D^\mathrm{eff}_{\mathrm{nf}}}{D_T} = 1 + \frac{Pe_s^2}{2\gamma^2}.
\end{equation}
For an oscillatory Poiseuille flow, $D^\mathrm{eff}$ after the initial transients becomes a periodic function of time. At long times, we define the time-averaged effective dispersion coefficient as 
\begin{equation}
\label{Deff_diffusion}
   \langle D^\mathrm{eff*} \rangle  = \lim_{t^\prime \to \infty}\frac{1}{T}\int_{t^{\prime}}^{t^{\prime} + T} D^\mathrm{eff*}(t^{*})\odiff t^{*},
\end{equation}
where $ T = 2\pi/\chi$ is the period of the flow oscillation. Similarly, one can define the time-averaged effective drift as $\langle U^{\mathrm{eff}*} \rangle$.

\section{Weak-swimming asymptotic analysis}
\label{Sec :Weakswimming}

In the weak-swimming limit,  characterized by $Pe_s \ll 1$, we pose regular expansions for the fields and transport coefficients:
\begin{eqnarray}
        g&=&g_0 +Pe_s\, g_1 +Pe_s^2\, g_2 +\cdots, \\
        b^*&=&b_0^* +Pe_s\, b_1^* +Pe_s^2\, b_2^* +\cdots, \\ 
        U^{\mathrm{eff}*}&=&U_0^{\mathrm{eff}*} +Pe_s\, U_1^{\mathrm{eff} *}+Pe_s^2\, U_2^{\mathrm{eff}*} +\cdots,\\  
        D^{\mathrm{eff}*}&=&D_0^{\mathrm{eff}*} +Pe_s\, D_1^{\mathrm{eff}*} +Pe_s^2\, D_2^{\mathrm{eff}*} +\cdots. 
\end{eqnarray}

\subsection{Passive Brownian particles}
\label{Subsec :Passive}
At $O(1)$, the particle is passive and the average field is given by  $g_0 \equiv 1/(2\pi)$. This means that the number density across the channel is uniform. As a result, the effective drift at $O(1)$ is given by $U_0^{\mathrm{eff}*}   = \overline{u^*}$,  which vanishes upon time-averaging.

\begin{figure}
    \centering
    \includegraphics[width=5in]{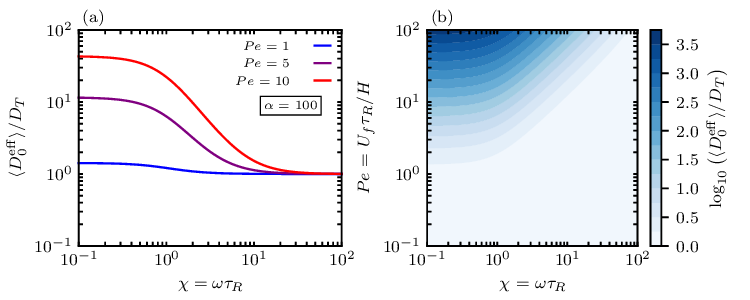}
    \caption{(a) Plots of the non-dimensional time-averaged effective dispersivity ($\langle D^{\mathrm{eff}}_{\mathrm{0}}\rangle/D_T$) as a function of $\chi$. (b) Contour plot of the logarithm of $\langle D^{\mathrm{eff}}_{\mathrm{0}}\rangle/D_T$ as a function of $Pe$ and $\chi$. For all results shown, $\alpha=100$, and $\gamma^2=0.1$.}
    \label{fig:passive}
\end{figure}

The displacement field at $O(1)$  admits a solution of the form $b_0^* = \Re[A_0^\prime(y^*) e^{i\chi t^*} /(2\pi)]$, where the solution to $A_0^\prime$ is provided in appendix \ref{app :passive-solution}. The instantaneous effective dispersion coefficient at $O(1)$ after initial transients is given by
\begin{equation}
\label{eq:D0eff}
    D_0^{\mathrm{eff}*}(t^*) = 1- \frac{1}{2\gamma^2}\int_{-1}^{1} u^{*}  \Re\left[A_0^\prime e^{i\chi t^{*}} \right]\odiff y^{*}.
\end{equation}
An analytical expression for the effective dispersion coefficient was derived by \citet{Watson_1983},  given by  
\begin{equation}
\label{eq: D0eff-analytical}
    \langle D^{\mathrm{eff}*}_0 \rangle = 1 +  
    \frac{Pe^2}{\kappa^2} \frac{\cosh{(2\kappa)} - \cos{(2\kappa)}}{\cosh{(2\kappa)} + \cos{(2\kappa)}}\frac{\iota(2\kappa) - \iota(\sqrt{2\chi}/\gamma)}{
    \chi^2 - 4\gamma^4\kappa^4},
\end{equation}
where
\begin{equation}
    \iota(a) = \frac{\sinh{(a)} -\sin{(a)}}{a\left(
    \cosh{(a)} - \cos{(a)}
    \right)}.
\end{equation}

In figure \ref{fig:passive}, we plot the passive dispersivity ($\left<D^{\mathrm{eff}}_{\mathrm{0}}\right>/D_T$), given in \eqref{eq: D0eff-analytical},  as a function of $\chi$ and $Pe$. Since $\alpha$ is held fixed, increasing $\chi$ corresponds to increasing the dimensional frequency. In the low frequency limit, we have  $\avg{D_0^{\mathrm{eff}}}/D_T \to 1 + 4Pe^2/(945\gamma^4)$ as $\chi \to 0$. For a steady Poiseuille flow of the same amplitude, the long-time dispersion coefficient $D_0^{\mathrm{eff}}/D_T=1 + 8Pe^2/(945\gamma^4)$. As is well known, in oscillatory flow,  $(\avg{D_0^{\mathrm{eff}}} - D_T)/D_T$ approaches half of its steady value as $\chi \to 0$ \citep{aris1960dispersion,bowden1965horizontal,van1982stochastic,Watson_1983,ng2006dispersion,chu2019dispersion,chu2020dispersion}. On the other hand, as $\chi\to \infty$, $\left<D^{\mathrm{eff}}_{\mathrm{0}}\right> /D_T  \to 1$ regardless of $Pe$ [see figure \ref{fig:passive}(a)]. In this high-frequency limit, shear-induced dispersion vanishes due to the rapid flow oscillations. For low and intermediate frequencies, $\left<D^{\mathrm{eff}}_{\mathrm{0}}\right>$ increases with $Pe$, as is consistent with Taylor dispersion. Overall, $\left<D^{\mathrm{eff}}_{\mathrm{0}}\right>$ decreases monotonically with increasing frequency until it reaches the high-frequency limit. In figure \ref{fig:passive}(b), we plot the same analytical expression given in equation \eqref{eq: D0eff-analytical} in a contour plot as a function of both $\chi$ and $Pe$.

\subsection{First order}
\label{Subsec :Firstorder}
At $O(Pe_s)$, the average field is governed by 
\begin{subequations}
\begin{equation}
        \frac{\partial g_1}{\partial t^*} + \frac{\partial }{\partial y^*}\left(  -\gamma^2 \frac{\partial g_1}{\partial y^*}\right) +\frac{\partial}{\partial \phi}\left( \Omega_f^* g_1 - \frac{\partial g_1}{\partial \phi} \right)=-q_y \frac{\partial g_0}{\partial y^*},
\end{equation}
\begin{equation}
     \gamma^2\frac{\partial g_1}{\partial y^*}=q_y g_0,\quad\mathrm{at}\quad y^*=\pm 1, 
\end{equation}
\begin{equation}
    \int_{-1}^1 \odiff y^*\intS g_1 \dq =0.
\end{equation}
\end{subequations}
Assuming a solution of the form $g_1 = A_1(y^*, t^*) \cos\phi + B_1(y^*,t^*)\sin\phi$, we obtain 
\begin{subequations}
\label{eq: O(1)-1}
    \begin{equation}
        \frac{\partial A_1}{\partial t^*} - \gamma^2 \frac{\partial^2 A_1}{\partial y^{*2}} + \Omega_f^* B_1 + A_1=0, 
    \end{equation}
    \begin{equation}
        \frac{\partial B_1}{\partial t^*} - \gamma^2\frac{\partial^2 B_1}{\partial y^{*2}} - \Omega_f^* A_1 +  B_1=0, 
    \end{equation}
    \begin{equation}
        \frac{\partial A_1}{\partial y^*}=0, \quad\mathrm{and}\quad \frac{\partial B_1}{\partial y^*}=\frac{1}{2\pi\gamma^2},
        \quad\mathrm{at}\quad y^* = \pm 1.
    \end{equation}
\end{subequations}
The instantaneous effective drift at this order $ U^{\mathrm{eff*}}_1$ vanishes.

The displacement field at $O(Pe_s)$ is governed by 
\begin{subequations}
\label{eq:b1-all}
\begin{align}
\label{eq:b1-eq}
    \frac{\partial b_1^*}{\partial t^*} + \frac{\partial }{\partial y^*}\left(  - \gamma^2\frac{\partial b_1^*}{\partial y^*}\right) +\frac{\partial }{\partial\phi}\left( \Omega^*_f b_1^* - \frac{\partial b_1^*}{\partial \phi}\right)&=
    -q_y \frac{\partial b_0^*}{\partial y^*}+\left(U_0^{\mathrm{eff}*} - u^*\right) g_1 \nonumber \\
    &\quad \, +\left(U_1^{\mathrm{eff}*} - q_x\right) g_0 ,
\end{align}
\begin{equation}
      \gamma^2 \frac{\partial b_1^*}{\partial y^*}=q_y b_0^*,\quad\mathrm{at}\quad y^*=\pm 1, 
\end{equation}
\begin{equation}
    \int_{-1}^1 \odiff y^*\intS b_1^* \dq =0,
\end{equation}
\end{subequations}
which admits a solution of the form $b_1^* = A_2(y^*, t^*) \cos\phi + B_2(y^*, t^*)  \sin\phi$. Inserting this form into equation~\eqref{eq:b1-all}, we obtain 
\begin{subequations}
\label{eq: O(1)-2}
    \begin{equation}
        \frac{\partial A_2}{\partial t^*} -  \gamma^2\frac{\partial^2 A_2}{\partial y^{*2}} + \Omega_f^* B_2 + A_2 = \left(U_0^{\mathrm{eff}*} - u^*\right) A_1 - g_0, 
    \end{equation}
    \begin{equation}
        \frac{\partial B_2}{\partial t^*} -  \gamma^2\frac{\partial^2 B_2}{\partial y^{*2}} - \Omega_f^* A_2 + B_2 =  - \frac{\partial b_0^*}{\partial y^*} + \left(U_0^{\mathrm{eff}*} - u^*\right) B_1, 
    \end{equation}
    \begin{equation}
        \frac{\partial A_2}{\partial y^*} =0,\quad\mathrm{and}\quad \frac{\partial B_2}{\partial y^*} = \frac{b_0^*}{ \gamma^2}, \quad\mathrm{at}\quad y^* = \pm 1. 
    \end{equation}
\end{subequations}
The effective longitudinal dispersivity at $O(Pe_s)$ vanishes, 
\begin{equation}
    D_1^{\mathrm{eff}*} = -\frac{1}{2 \gamma^2}\int_{-1}^{1}\odiff y^{*}\intS(u^{*}b_{1}^{*} + q_xb_0^{*})\dq = 0.
\end{equation}

\subsection{Second order}
\label{Subsec :secondorder}
At $O(Pe_s^2)$, the average field is governed by
\begin{subequations}
    \begin{equation}
    \label{eq:g_2diff}
    \frac{\partial g_2}{\partial t^{*}} + \frac{\partial}{\partial y^{*}}
    \left( q_yg_1 - \gamma^2 \frac{\partial g_2}{\partial y^*}  
    \right) + \frac{\partial}{\partial \phi}\left(\Omega_f^* g_2 - \frac{\partial g_2}{\partial \phi}
    \right) = 0,
\end{equation}
\begin{equation}
    \gamma^2 \frac{\partial g_2}{\partial y^*} = q_yg_1 \quad \text{at} \quad y^* = \pm 1,
\end{equation}
\begin{equation}
    \int_{-1}^{1} \odiff y^* \intS g_2\dq= 0.
\end{equation}
\end{subequations}
We propose a solution of the form,
\begin{equation}
\label{eq:g_2}
 g_2 = K_1(y^*,t^*) + C_1(y^*,t^*)\cos{2\phi} + D_1(y^*,t^*)\sin{2\phi}.   
\end{equation}

The displacement filed at  $O(Pe_s^2)$ is governed by
\begin{subequations}
    \begin{multline}
    \label{eq:b_2diff}
        \frac{\partial b_2^*}{\partial t^*} + \frac{\partial}{\partial y^*}\left(
        q_yb_1^* - \gamma^2 \frac{\partial b_2^*}{\partial y^*}
        \right) + \frac{\partial}{\partial \phi}\left(\Omega_f^*b_2^* - \frac{\partial b_2^*}{\partial \phi}\right) \\
        =  \left( U_0^{\text{eff*}} - u^*\right)g_2 + \left(U_1^{\text{eff*}} - q_x  \right)g_1 + U_2^{\text{eff*}}g_0,
    \end{multline}
    \begin{equation}
        \gamma^2\frac{\partial b_2^*}{\partial y^*} = q_yb_1^* \quad \text{at} \quad y^* = \pm 1,
    \end{equation}
    \begin{equation}
        \int_{-1}^{1}\odiff y^*\intS b_2^* \dq = 0.
    \end{equation}
\end{subequations}

We assume a solution for $b_2^*$, 
\begin{equation}
\label{eq:b_2}
    b_2^* = K_2(y^*,t^*) + C_2(y^*,t^*)\cos{2\phi} + D_2(y^*,t^*)\sin{2\phi}.
\end{equation}
One can show that $\langle U_2^{\mathrm{eff}*} \rangle =0$, and 
\begin{equation}
\label{eq: D2eff}
    D_2^{\mathrm{eff}*} = -\frac{1}{2\gamma^2}\int_{-1}^{1}\odiff y^{*}\intS(u^*b_2^{*} + q_xb_1^{*})\dq =-\frac{\pi}{2\gamma^2}\int_{-1}^1 \left(2 u^* K_2 + A_2  \right)\odiff y^* .
\end{equation}
To obtain $D_2^{\mathrm{eff}*}$, one needs to solve for $K_2$. The relevant equations are given by

\begin{subequations}
\label{eq: O(2)}
    \begin{equation}
        \frac{\partial K_1}{\partial t^*} - \gamma^2\frac{\partial^2 K_1}{\partial y^{*2}} + \frac{1}{2}\frac{\partial B_1}{\partial y^{*}} = 0,
    \end{equation}
    \begin{equation}
        \frac{\partial K_2}{\partial t^*} + \left[
        \frac{1}{2}\frac{\partial B_2}{\partial y^{*}} - \gamma^2\frac{\partial^2K_2}{\partial y^{*2}}
        \right] = U_2^{\mathrm{eff*}}g_0 - \frac{1}{2}A_1 + \left(
        U_0^{\mathrm{eff*}} - u^{*}
        \right)K_1,
    \end{equation}
    \begin{equation}
        \frac{\partial K_1}{\partial y^{*}} = \frac{1}{2\gamma^2} B_1, \quad \mathrm{and} \quad \frac{\partial K_2}{\partial y^{*}} = \frac{1}{2\gamma^2}B_2 \quad \mathrm{at} \quad y^{*} = \pm 1.
    \end{equation}
\end{subequations}

We solve equations \eqref{eq: O(1)-1}, \eqref{eq: O(1)-2} and \eqref{eq: O(2)} using a Chebyshev collocation method. For time evolution, we use the Crank-Nicolson method. At long times, the time-averaged dispersion coefficient,  $\langle D^{\mathrm{eff*}}_2 \rangle$, is obtained via numerical integration over one oscillation period. We also solve the full GTD theory by solving equations \eqref{eq:g-non-dimensional} and \eqref{eq:b-non-dimensional} numerically (see appendix D). To extract an approximation of $D_2^\mathrm{eff*}$ from the full solution, denoted as $\tilde{D}_2^\mathrm{eff*}$, we use the relation $\langle \tilde{D}_2^\mathrm{eff*} \rangle=\left( \langle D^\mathrm{eff*}\rangle - \langle D_0^\mathrm{eff*} \rangle\right)/Pe_s^2 $. Here, $\langle D_0^\mathrm{eff*} \rangle$ is the analytical solution for passive particles from equation \eqref{eq: D0eff-analytical}, and the full simulation is performed with $Pe_s=0.1$.

\begin{figure}
    \centering
    \includegraphics[width=5in]{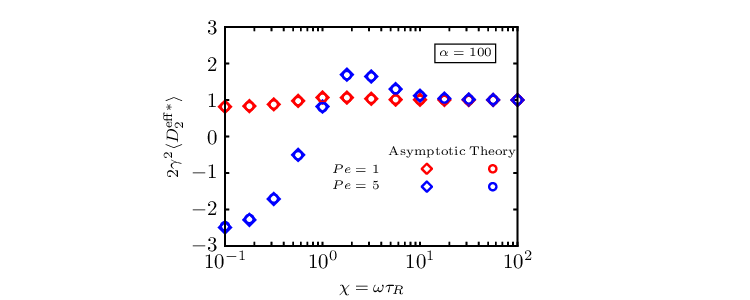}
    \caption{The $O(Pe_s^2)$ dispersivity as a function of $\chi$. For all results, $\alpha=100$, and $\gamma^2=0.1$. Circles denote results obtained from the numerical solutions of the full GTD theory for $Pe_s=0.1$.
    Diamonds denote results from the asymptotic analysis.
    }
    \label{fig:D2eff}
\end{figure}

In figure \ref{fig:D2eff}, we plot $\langle D^{\mathrm{eff*}}_2\rangle$ as a function of $\chi$. The asymptotic results (diamonds) are compared with numerical solutions (circles) of the full GTD theory (see appendix \ref{app :Numerical}). As in the passive case (see figure \ref{fig:passive}), the shear-induced dispersion vanishes in the high-frequency limit. From \eqref{eq: noflowdispersivity}, we have $\langle D^\mathrm{eff*}_2 \rangle\to 1/(2\gamma^2) $ as $\chi \to \infty$. Indeed, figure~\ref{fig:D2eff} shows that $2 \gamma^2\, \langle D^{\mathrm{eff*}}_2 \rangle$ approaches unity in the high-frequency limit.

Overall,  $\langle D^{\mathrm{eff*}}_2\rangle$ can be either positive or negative depending on $Pe$ and $\chi$. This means that activity can either enhance or hinder the longitudinal dispersion in an oscillatory flow compared to the passive case. In particular, a reduction in the dispersion ($\langle D^{\mathrm{eff*}}_2\rangle<0$) occurs in the low-frequency regime  when $Pe$ is sufficiently large (e.g., $Pe=5$; blue markers). This reduction can be attributed to shear-reduced swim diffusion \citep{peng2020upstream}, which becomes prominent for sufficiently strong shear. For $Pe=1$, $\langle D^{\mathrm{eff*}}_2\rangle>0$ for all values of $\chi$.  For $Pe=5$, $\langle D^{\mathrm{eff*}}_2\rangle$ can be either positive or negative depending on $\chi$. There exists an optimal frequency at which the enhancement in dispersion is maximized.

    To understand this effect in association with shear-reduced swim diffusion, recall that the effective dispersion coefficient given by \eqref{eq:Deff} consists of three contributions: translational diffusion $D_T$, shear-modified swim diffusion $-U_s \overline{\tilde{m}_x}$, and classical Taylor dispersion $-\overline{u\tilde{n}}$. The latter two contributions are coupled and therefore should not be interpreted independently as diffusivities. As $Pe$ increases, the fluid vorticity increases proportionally, reducing the effective run length of active particles and thereby suppressing the swim diffusion contribution. In contrast, the Taylor dispersion term increases and becomes significant at larger $Pe$. As a result, only in the intermediate-$Pe$ regime is the net dispersion reduced compared with the case without flow.

\begin{figure}
    \centering
    \includegraphics[width=5in]{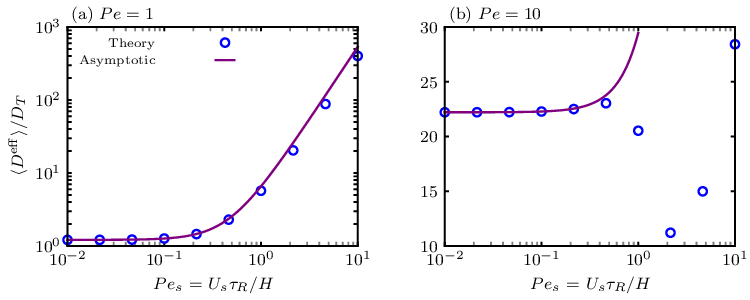}
    \caption{ Plots of $\langle D^{\mathrm{eff}}\rangle/D_T$ as a function of $Pe_s$ for (a) $Pe=1$, and (b) $Pe=10$. The solid lines denote the two-term asymptotic solution, $\langle D_0^{\mathrm{eff*}}\rangle + Pe_s^2\langle D_2^{\mathrm{eff*}}\rangle$. Circles are numerical solutions of the full GTD theory. For all results shown, $\chi=1$, $\gamma^2=0.1$, and $\kappa=0.1$. } 
    \label{fig:Asymptote}
\end{figure}

In figure \ref{fig:Asymptote}, we compare $\langle D^{\mathrm{eff}}\rangle/D_T$ from the two-term asymptotic solution (solid lines), $\langle D_0^{\mathrm{eff*}} \rangle + Pe_s^2 \langle D_2^{\mathrm{eff*}} \rangle$, with the numerical solutions (circles) of the full GTD theory. In figure \ref{fig:Asymptote}(a), for $Pe = 1$, the two-term asymptotic solution (solid line) agrees with the full GTD theory (circles) well beyond its formal regime of validity, i.e., $Pe_s \ll 1$. In figure \ref{fig:Asymptote}(b), for a stronger flow ($Pe = 10$), the full GTD results (circles) show that $\langle D^{\mathrm{eff}} \rangle / D_T$ varies non-monotonically with increasing $Pe_s$. 
As $Pe_s$ increases beyond the weak-swimming regime, the effective dispersivity decreases due to shear-reduced swim diffusion. The effective dispersivity increases again when activity ($Pe_s$) is sufficiently high. This behavior is not captured by the asymptotic solution (solid line), which is valid only in the weak-swimming limit. 

The above weak-swimming analysis applies to microorganisms in wide channels with moderate run lengths. For \textit{E. coli}, $U_s\sim 30~ \mu$m/s \citep{turner2000real}, $\tau_R\sim 1~$s. In a channel of half-width $H \sim 100~\mu$m, this gives $Pe_s \sim 0.3$. For active particles that have larger run lengths, or in  narrower channels, it is then necessary  to consider the finite activity regime. 

\section{Dispersivity in the finite activity regime}
\label{Sec :Results}
To characterize the general behavior of the effective dispersion coefficient, we resort to numerical solutions of the full GTD theory (see appendix \ref{app :Numerical}). The GTD equations are evolved over time.  Numerical solutions of the GTD theory are compared to results obtained from BD simulations (see appendix \ref{app :BD}).

\subsection{Competition between flow advection and particle activity} 
\label{Subsec :flowadvection}

\begin{figure}
    \centering
    \includegraphics[width = 5in]{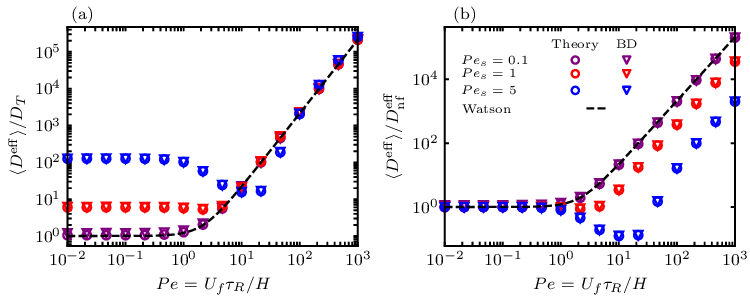}
    \caption{(a) Plots of $\langle D^{\mathrm{eff}} \rangle/ D_T$ as a function of $Pe$ for several values of $Pe_s$.
    (b) Plots of  $\langle D^{\mathrm{eff}} \rangle/ D^{\mathrm{eff}}_{\mathrm{nf}}$ as a function of $Pe$ for several values of $Pe_s$. Circles represent solutions of the full GTD theory, and triangles denote results from BD simulations. The dashed line represents the passive ($Pe_s = 0$) results. For all results, $\chi = 1, \gamma^2 = 0.1$, and $\kappa = 0.1$.
    }
    \label{fig:Deff/DT}
\end{figure}

In this section, we examine the dispersion behavior of ABPs for a given flow oscillation frequency, $\chi=1$. With this fixed frequency, the dispersion is qualitatively similar to that considered by \citet{peng2020upstream} for a steady Poiseuille flow. In figure \ref{fig:Deff/DT}(a), we plot $\langle D^{\mathrm{eff}} \rangle / D_T$ as a function of $Pe$ for different values of $Pe_s$. As $Pe \to 0$, we recover the dispersion coefficient in the absence of flow, $D^{\mathrm{eff}}_{\mathrm{nf}}$. Since $D^{\mathrm{eff}}_{\mathrm{nf}}/D_T = 1+Pe_s^2/(2\gamma^2)$, the low-$Pe$ plateau in the dispersion coefficient increases with activity ($Pe_s$). On the other hand, as $Pe \to \infty$, the flow speed dominates over the swim speed. In this regime, the effective dispersion coefficient converges to the passive result (dashed line), regardless of activity. For higher activity (e.g., $Pe_s=5$; blue markers), a large flow amplitude ($Pe$) is required for the dispersion coefficient to approach the passive result. For $Pe_s = 0.1$ (purple markers) and $Pe_s = 1$ (red markers), the swimming effects are largely dominated by the Taylor dispersion component. When activity is sufficiently high (e.g., $Pe_s=5$; blue markers), the dispersion coefficient varies non-monotonically with increasing flow amplitude.  The reduction in $\avg{D^\mathrm{eff}}$ for intermediate flow amplitudes are due to the shear-reduced swim diffusion (see also \S~\ref{Subsec :secondorder}). 

In figure \ref{fig:Deff/DT}(b), we replot the data shown in figure \ref{fig:Deff/DT}(a) using a different scaling---$\langle D^{\mathrm{eff}} \rangle / D_\mathrm{nf}^\mathrm{eff}$ instead of $\langle D^{\mathrm{eff}} \rangle / D_T$. By scaling the effective dispersion with the no-flow dispersion coefficient, all curves collapse in the low-$Pe$ limit. Conversely, the rescaled dispersion coefficient approaches different values in the large-$Pe$ limit.

\begin{figure}
    \centering
    \includegraphics[width=5in]{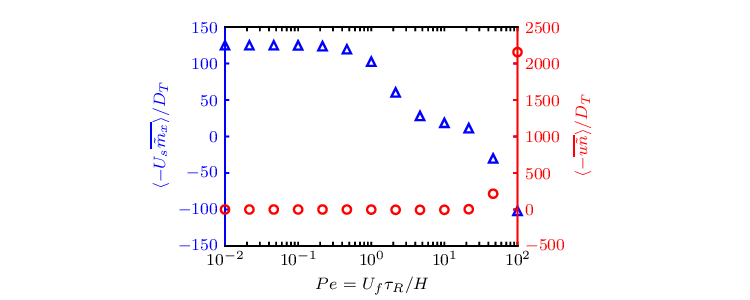}
    \caption{Plots of the two contributions to $\langle D^{\mathrm{eff}} \rangle / D_T$, $\langle -U_s \overline{\tilde{m}_x} \rangle / D_T$ and $\langle - \overline{u\tilde{n}} \rangle / D_T$ as a function of $Pe$ for $Pe_s = 5$.
    Blue triangles represent $\langle -U_s \overline{\tilde{m}_x} \rangle / D_T$, and red circles represent $\langle - \overline{u\tilde{n}} \rangle / D_T$.
    All results are obtained by solving the full GTD theory with $\chi = 1$, $\gamma^2 = 0.1$, and $\kappa = 0.1$.}
    \label{fig:Deffsplit}
\end{figure}

To visualize the non-monotonic behavior of $\langle D^{\mathrm{eff}} \rangle / D_T$, we plot its two contributions, $\langle-U_s\overline{\tilde{m}_x}\rangle/D_T$ and $\langle -\overline{u\tilde{n}}\rangle /D_T$ as functions of $Pe$ in figure \ref{fig:Deffsplit}. In the intermediate $Pe$ regime, $\langle-U_s\overline{\tilde{m}_x}\rangle/D_T$ decreases with increasing $Pe$ and even becomes negative at high $Pe$. In contrast, $\langle -\overline{u\tilde{n}}\rangle/D_T$ increases with $Pe$. For sufficiently large $Pe$, the Taylor component dominates over the swim contribution. The competition between these two contributions gives rise to the observed non-monotonicity in the effective dispersivity, as shown in figure \ref{fig:Deff/DT}. We emphasize that these two contributions are not independent; therefore, each individual term should not be interpreted as a dispersion coefficient.

\subsection{Effect of oscillation frequency} \label{Subsec :oscillation frequency}

\begin{figure}
    \centering
    \includegraphics[width=5in]{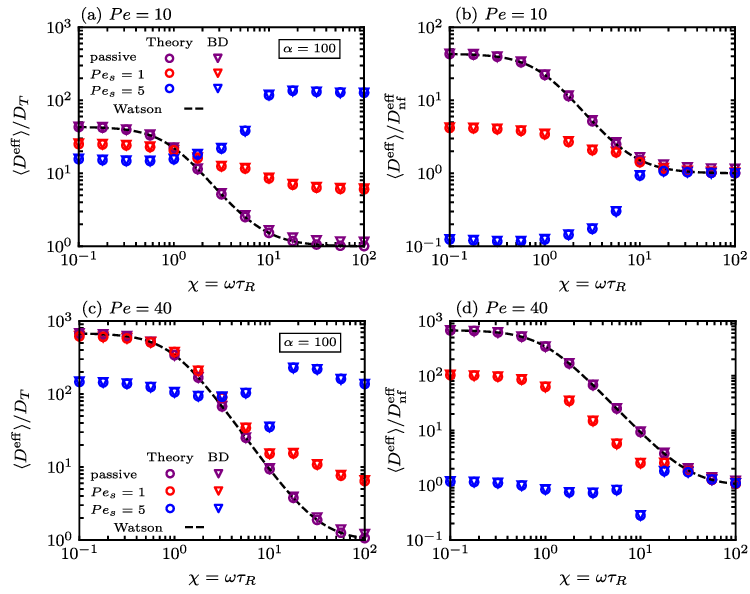}
    \caption{Plots of $\langle D^{\mathrm{eff}}\rangle/D_T $ as a function of $\chi$ for different values of $Pe_s$, shown for (a) $Pe = 10$ and (c) $Pe = 40$. Plots of $\langle D^{\mathrm{eff}}\rangle/ D^{\mathrm{eff}}_{\mathrm{nf}}$ as a function $\chi$ for different values of $Pe_s$, shown for (b) $Pe = 10$, and (d) $Pe = 40$. For all results shown, $\alpha = 100$, and $\gamma^2 = 0.1$. Circles denote results obtained from numerical solutions of the full GTD theory, and triangles represent results from BD simulations. The dashed line indicates the passive ($Pe_s = 0$) solution of \citet{Watson_1983}.}
    \label{fig:Deff_fixPe_diffPes_chi}
\end{figure}

We now consider the effect of flow oscillation frequency on the effective longitudinal dispersion. In figure \ref{fig:Deff_fixPe_diffPes_chi}, we plot $\langle D^{\mathrm{eff}}\rangle/D_T$ and $\langle D^{\mathrm{eff}}\rangle/D^{\mathrm{eff}}_{\mathrm{nf}}$ as functions of $\chi$ for different values of $Pe$ and $Pe_s$.  Circles denote numerical solutions of the GTD theory, while triangles represent results from BD simulations.  The dashed line corresponds to the passive case (i.e., $Pe_s = 0$), as determined by \citet{Watson_1983}. Figures \ref{fig:Deff_fixPe_diffPes_chi}(a) and \ref{fig:Deff_fixPe_diffPes_chi}(b) correspond to $Pe = 10$, whereas figures \ref{fig:Deff_fixPe_diffPes_chi}(c) and \ref{fig:Deff_fixPe_diffPes_chi}(d) correspond to  $Pe = 40$. To isolate the effect of the oscillation frequency, we fix $\alpha$ in this section. With $\alpha$ fixed, we note that $\kappa$ depends explicitly on $\chi$ (i.e., $\kappa^2 = \chi/\alpha$).

In the high-frequency limit, one can show that the flow vanishes at leading order (see appendix \ref{app :high-frequency} for the asymptotic analysis). Effectively, the dispersion in the high-frequency limit is equivalent to the no-flow case. Therefore, $\avg{D^\mathrm{eff}}\to D_\mathrm{nf}^\mathrm{eff}$ as $\chi \to \infty$,  regardless of $Pe_s$ or $Pe$.   Recall that,  in the absence of flow, $\avg{D^\mathrm{eff}} = D_T+ D^\mathrm{swim} = D_T +U_s^2\tau_R/2$, or equivalently $\avg{D^\mathrm{eff}}/D_T = 1+Pe_s^2/(2\gamma^2)$. As shown in figures \ref{fig:Deff_fixPe_diffPes_chi}(a) and \ref{fig:Deff_fixPe_diffPes_chi}(c),  in this large-$\chi$ limit this leads to larger values of $\avg{D^\mathrm{eff}}/D_T$ for increasing $Pe_s$. Upon scaling the effective dispersion by $D^{\mathrm{eff}}_{\mathrm{nf}}$, the ratio $\langle D^{\mathrm{eff}}\rangle/D^{\mathrm{eff}}_{\mathrm{nf}}$ approaches unity in the high-frequency regime, as shown in figures \ref{fig:Deff_fixPe_diffPes_chi}(b) and \ref{fig:Deff_fixPe_diffPes_chi}(d).
Compared to the cases shown in figures \ref{fig:Deff_fixPe_diffPes_chi}(a) and \ref{fig:Deff_fixPe_diffPes_chi}(b), a higher value of $\chi$ is required to reach the high-frequency limit in figures \ref{fig:Deff_fixPe_diffPes_chi}(c) and \ref{fig:Deff_fixPe_diffPes_chi}(d), owing to their larger flow amplitudes ($Pe$). 
    In figures \ref{fig:Deff_fixPe_diffPes_chi}(a) and \ref{fig:Deff_fixPe_diffPes_chi}(c), in the low-frequency regime, $\avg{D^{\mathrm{eff}}}/D_T$ is lower for higher $Pe_s$,  due to the non-monotonic dependence of dispersion on $Pe$ as discussed in  figure \ref{fig:Deff/DT}(a) [see also  \citet{peng2020upstream}]. For $Pe = 10$ and $40$, this regime lies in the range where shear suppresses swim diffusion, while classical Taylor dispersion has not yet become dominant.  Consequently, particles with higher swim speeds (e.g., $Pe_s=5$ versus $Pe_s=1$) loses a larger fraction of their swim diffusivity. 
We note that $D_\mathrm{nf}^\mathrm{eff}$ depends explicitly on $Pe_s$. In the low-frequency regime, $\langle D^{\mathrm{eff}}\rangle/D^{\mathrm{eff}}_{\mathrm{nf}}$ is even lower than $\avg{D^{\mathrm{eff}}}/D_T$ for higher $Pe_s$ because $D^{\mathrm{eff}}_{\mathrm{nf}}$ increases quadratically with the swim speed, whereas $\langle D^{\mathrm{eff}}\rangle$ does not increase as rapidly (see figure \ref{fig:Deff/DT}).

As shown in figure \ref{fig:Deff_fixPe_diffPes_chi},  the scaled dispersion coefficient $\avg{D^\mathrm{eff}}/D_\mathrm{nf}^\mathrm{eff}$ for passive particles decreases monotonically with $\chi$. For active particles, the scaled dispersion coefficient exhibits rich behavior that depends on the flow ($Pe$) and swim ($Pe_s$) speeds. At low activity [e.g., $Pe_s=1$ in figure \ref{fig:Deff_fixPe_diffPes_chi}(b)], the scaled dispersion coefficient remains a monotonically decreasing function of $\chi$. For higher activity [e.g., $Pe_s=5$ in figure \ref{fig:Deff_fixPe_diffPes_chi}(b)], the scaled dispersion coefficient begins at a low plateau and increases as a function of $\chi$, eventually converging to the common high-frequency limit. 

In figures \ref{fig:Deff_fixPe_diffPes_chi}(c) and \ref{fig:Deff_fixPe_diffPes_chi}(d), at low activity ($Pe_s = 1$; red markers), the effective dispersion coefficient $\avg{D^{\mathrm{eff}}}/D_T$ and the scaled dispersion coefficient $\avg{D^\mathrm{eff}}/D_\mathrm{nf}^\mathrm{eff}$ decrease monotonically with $\chi$, exhibiting a plateau around $\chi \sim 10$ and then continuing to decrease towards the no-flow limit. However, in figure \ref{fig:Deff_fixPe_diffPes_chi}(d), the scaled dispersion coefficient exhibits oscillatory behavior as a function of $\chi$ for $Pe_s=5$. This non-trivial variation occurs when the flow P\'{e}clet number is sufficiently large. For $Pe_s=5$ and $Pe=40$, we see that $\avg{D^\mathrm{eff}}/D_\mathrm{nf}^\mathrm{eff}$ exhibits both a minimum and a maximum at different frequencies. The observed oscillatory behavior likely results from resonance in which the flow oscillation timescale matches an intrinsic timescale of the ABPs. Because the dynamics of ABPs involve multiple timescales, the intrinsic timescale that leads to resonant diffusion cannot be easily obtained. In the following, we investigate this phenomenon numerically by identifying the thresholds of this oscillation as a function of $\chi$, $Pe$, and $Pe_s$. Physically, $\chi$, $Pe$, and $Pe_s$ characterize the flow oscillation timescale $1/\omega$, the flow advection timescale $H/U_f$, and the swimming timescale $H/U_s$, respectively.

\begin{figure}
    \centering
    \includegraphics[width=5in]{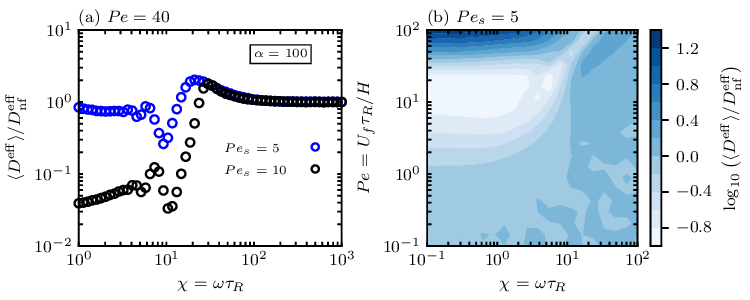}
    \caption{(a) Plots of $\langle D^{\mathrm{eff}}\rangle/D^{\mathrm{eff}}_{\mathrm{nf}}$ as a function of $\chi$ for different values of $Pe_s$. (b) Contour plot of the logarithm of $\langle D^{\mathrm{eff}}\rangle/D^{\mathrm{eff}}_{\mathrm{nf}}$ as a function of $Pe$ and $\chi$ at $Pe_s = 5$. All results are from BD simulations with $\alpha = 100$, and $\gamma^2 = 0.1$. The contour plot is produced from a total of 400 data points, with 20 points uniformly spaced in logarithmic space along each axis. }
    \label{fig:contourPechi}
\end{figure}

We first examine the details of the oscillation in figure \ref{fig:contourPechi}(a) by plotting $\langle D^{\mathrm{eff}}\rangle/D^{\mathrm{eff}}_{\mathrm{nf}}$ as a function of $\chi$ for two values of $Pe_s$, using more data points than in figure \ref{fig:Deff_fixPe_diffPes_chi}. Even though it is not straightforward to determine the intrinsic timescale associated with resonant diffusion, one can rationalize its variation as a function of the swim speed. Compared to $Pe_s = 5$ (blue circles), the onset of oscillatory behavior of $\langle D^{\mathrm{eff}}\rangle/D^{\mathrm{eff}}_{\mathrm{nf}}$ and the locations of its extrema shift to higher flow frequencies for higher activity ($Pe_s = 10$, black circles). This can be understood by considering the swim timescale, $\tau_s=H/U_s$, which characterizes the time it takes for the ABPs to traverse the channel in the transverse direction. As the swim speed ($Pe_s$) increases, $\tau_s$ decreases. Therefore, a smaller flow oscillation timescale (or higher flow oscillation frequency) is required to match the swim timescale.

Next, we consider how the oscillation in the dispersion coefficient further depends on the flow advection. In figure \ref{fig:contourPechi}(b), we show a contour plot of the logarithm of $\langle D^{\mathrm{eff}}\rangle/D^{\mathrm{eff}}_{\mathrm{nf}}$ as a function of $Pe$ and $\chi$ for $Pe_s = 5$. We observe that as $Pe$ increases, the extrema in $\langle D^{\mathrm{eff}}\rangle/D^{\mathrm{eff}}_{\mathrm{nf}}$ shift to higher flow oscillation frequencies, reflected in the upward shift of the light-colored region in figure \ref{fig:contourPechi}(b) with increasing $\chi$. In addition to the swim timescale discussed in figure \ref{fig:contourPechi}(a), the ABPs in the presence of flow also have a flow timescale defined by $H/U_f$. As $Pe$ increases, this flow timescale decreases. To achieve resonance, the timescale defined by the flow oscillation, $1/\omega$, needs to be smaller. Due to the interplay of these timescales, in general $\langle D^{\mathrm{eff}}\rangle/D^{\mathrm{eff}}_{\mathrm{nf}}$ exhibits non-monotonic behaviors as a function of $Pe$ and $\chi$. Furthermore, we note that the region where the scaled dispersion coefficient attains low values shrinks as $Pe$ increases.

Together with the behavior of $\langle D^{\mathrm{eff}} \rangle / D^{\mathrm{eff}}_{\mathrm{nf}}$ shown in figure \ref{fig:contourPechi}(a), we note that increasing either $Pe$ or $Pe_s$ shifts the onset of oscillatory behavior to higher flow frequencies. This suggests that resonance can occur when the flow oscillation timescale matches some intrinsic timescale that results from the coupling between the swimming motion and the oscillatory flow advection. Formally, we have $\tau = \tau(Pe_s, Pe)$, where $\tau$ is the timescale that must match the flow oscillation timescale to achieve resonance. Extracting the functional dependence of $\tau$ on $H/U_s$ and $H/U_f$ is not pursued here. We note that resonant diffusion is commonly observed in both passive and active particle systems \citep{Castiglione_1998,leahy2015effect,khatri2022diffusion,chepizhko2022resonant}.

\subsection{Non-spherical particles}
\label{Subsec :Non-spherical}

 \begin{figure} 
    \centering
    \includegraphics[width=5in]{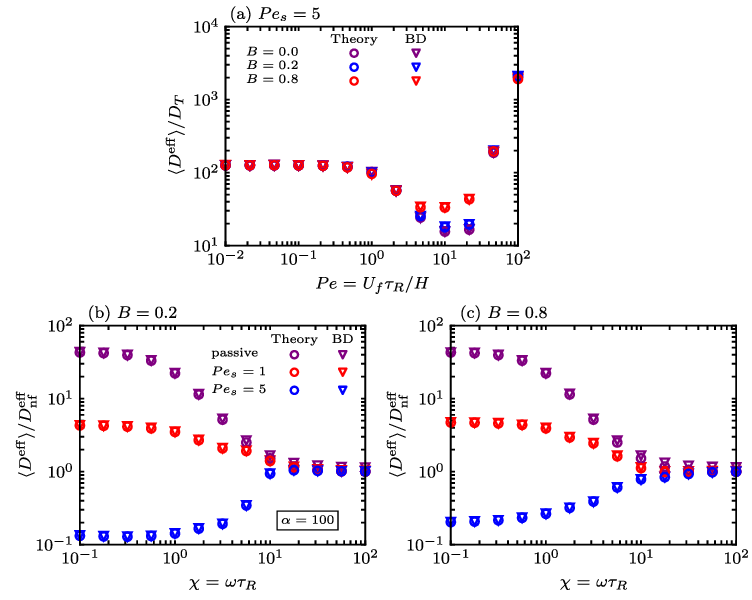}
    \caption{(a) Plots of $\langle D^{\mathrm{eff}} \rangle / D_T$ as a function of $Pe$ for different values of $B$. For all results in (a), $Pe_s = 5$, $\chi = 1$, $\gamma^2 = 0.1$, and $\kappa = 0.1$. (b) Plots of $\langle D^{\mathrm{eff}}\rangle/D^{\mathrm{eff}}_{\mathrm{nf}}$ as a function of $\chi$  for $B = 0.2$. (c) Plots of $\langle D^{\mathrm{eff}}\rangle/D^{\mathrm{eff}}_{\mathrm{nf}}$ as a function of $\chi$  for $B = 0.8$.  For all results shown, circles represent results from numerical solutions of the full GTD theory, while triangles denote results from BD simulations. The labels shown in (b) also apply to the corresponding curves in (c). For (b) and (c), $Pe = 10, \alpha = 100$, and $\gamma^2 = 0.1$. }
    \label{fig:non-spherical} 
\end{figure}

We now analyze how the effective dispersivity is influenced by the shape of the particle. For an ellipsoidal particle, a shape factor is defined as $B = (r^2 - 1)/(r^2 + 1)$, where $r = a/b$. Here, $a$ and $b$ denote the lengths of the semi-major and semi-minor axes, respectively. For a sphere, $r=1$ and $B=0$. For a thin rod, we have $B \to 1$ as $r\to \infty$. Modeling the angular dynamics using Jeffery equation \citep{Jeffery1922TheMO}, we have $\bOmega_f=\frac{1}{2}\nabla \times \bu_f + B \bq\times \left( \bm{E}\cdot\bq\right) $, where $E_{ij} =\frac{1}{2}\left( \frac{\partial u_i}{\partial x_j} + \frac{\partial u_j}{\partial x_i}\right) $ is the rate-of-strain tensor. In the oscillatory Poiseuille flow, we have 
\begin{equation} \label{eq:Omega_nonspherical}
    \Omega^{\prime *} = \Omega^\prime \tau_D =\frac{(1-i) Pe}{2\kappa} \left( 1 - B \cos2\phi \right) \sinh\left( (1+i)\kappa y^* \right) \sech\left((1+i)\kappa \right).
\end{equation}
As in the spherical case, we assume a constant translational diffusivity and enforce the no-flux boundary condition \eqref{eq:bc-noflux}.

In figure \ref{fig:non-spherical}(a), we plot $\langle D^{\mathrm{eff} }\rangle/D_T$ as a function $Pe$ for different values of $B$. The vorticity term ($\nabla \times \bu_f$) in the angular velocity induces spinning on ABPs, which reduces their persistence and consequently their swim diffusion. For non-spherical particles ($B \neq 0$), the additional alignment term from the rate-of-strain tensor is present. Because of this alignment, non-spherical particles lose less of their persistence. As a result, we observe that in figure \ref{fig:non-spherical}(a), the minimum in the dispersion coefficient decreases as $B$ decreases. As shown in figures \ref{fig:non-spherical}(b) and \ref{fig:non-spherical}(c), the scaled dispersion coefficient  $\langle D^{\mathrm{eff}}\rangle/D^{\mathrm{eff}}_{\mathrm{nf}}$  exhibits qualitatively similar behavior to that of spherical particles.  However, the suppression in effective dispersivity is reduced, owing to the reduced spinning.

In the current model, we have assumed an isotropic translational diffusivity for non-spherical particles and applied the same no-flux boundary conditions as for spherical particles \citep{ezhilan2015transport}. The anisotropic translational diffusivity of a spheroidal particle can be incorporated into the GTD framework \citep{kumar2021taylor, khair2022taylor}. Additional dynamics, such as alignment of a non-spherical particle with the channel wall, may also be included; such effects are expected to modify the dispersion behavior of non-spherical particles more significantly.

\subsection{Effect of $\alpha$}
\label{subsec:alpha}

\begin{figure}
    \centering
    \includegraphics[width=5in]{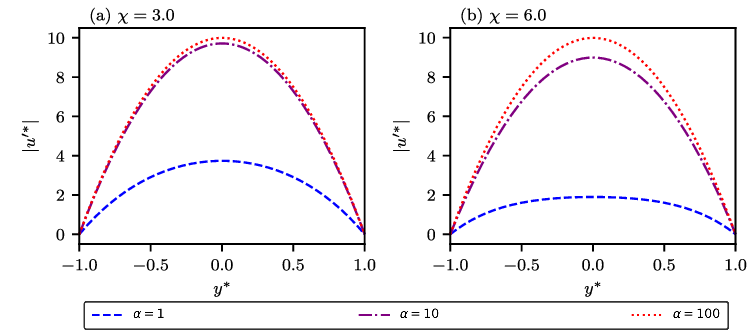}
    \caption{  
       (a) Plots of the norm of $u^{\prime *}$ as a function of $y^{*}$ for  $\chi = 3$.  (b) Plots of the norm of $u^{\prime *}$ as a function of $y^{*}$ for  $\chi = 6$. For all results shown, $Pe = 10$.
      } 
    \label{fig:flowfield}
\end{figure}

In the results presented so far, we have fixed $\alpha=100$.  We now examine the variation of the effective dispersion coefficient with decreasing $\alpha$ for spherical particles. For a fixed $\chi$ and channel geometry, the viscous length $1/\lambda = H\sqrt{\alpha/\chi} $ becomes smaller as $\alpha$ decreases. Consequentially, the flow strength is reduced as $\alpha$ decreases. To visualize the effect of the viscous length on the flow, in figure \ref{fig:flowfield} we plot the norm of $u^{\prime *}$ [see equation \eqref{eq: nondimlinvel}], as a function of $y^{*}$ for different values of $\alpha$. As shown in figure \ref{fig:flowfield}, for a given $\chi$, the amplitude of the flow decreases as $\alpha$ is reduced. 

\begin{figure}
    \centering
    \includegraphics[width=5in]{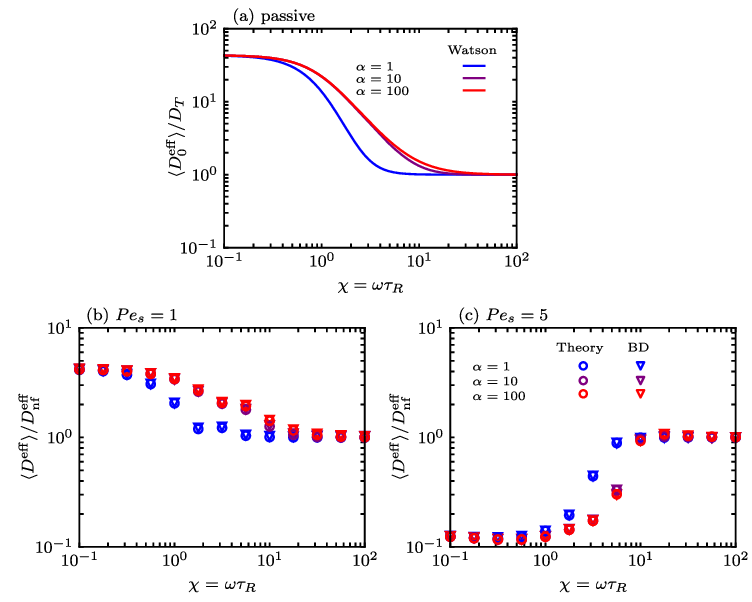}
    \caption{ (a) Plots of $\avg{D^{\mathrm{eff}}}/D^{\mathrm{eff}}_{\mathrm{nf}}$ as a function of $\chi$ for different values of $\alpha$ for passive ($Pe_s = 0$) particles. In the absence of activity, $\avg{D^{\mathrm{eff}}}/D^{\mathrm{eff}}_{\mathrm{nf}} = \avg{D_0^{\mathrm{eff}}}/D_T$. (b) Plots of $\avg{D^{\mathrm{eff}}}/D^{\mathrm{eff}}_{\mathrm{nf}}$ as a function of $\chi$ for different values of $\alpha$ for  $Pe_s = 1$. (c) Plots of $\avg{D^{\mathrm{eff}}}/D^{\mathrm{eff}}_{\mathrm{nf}}$ as a function of $\chi$ for different values of $\alpha$ for  $Pe_s = 5$. For all results shown in (b) and (c), circles represent results from numerical solutions of the full GTD theory, while triangles denote results from BD simulations. For all results shown, $Pe = 10$ and $\gamma^2 = 0.1$.   }
    \label{fig:alpha}
\end{figure}

In figure \ref{fig:alpha}, we plot $\avg{D^{\mathrm{eff}}}/D^{\mathrm{eff}}_{\mathrm{nf}}$ as a function of $\chi$ for different $Pe_s$ and $\alpha$. For passive particles, $D^{\mathrm{eff}}_{\mathrm{nf}} = D_T$ and $\avg{D^{\mathrm{eff}}} = \avg{D_0^{\mathrm{eff}}}$, where the subscript $0$ denotes $Pe_s=0$ (see \S~\ref{Sec :Weakswimming}). The passive results of \citet{Watson_1983} are plotted in figure \ref{fig:alpha}(a). Because the flow magnitude is weaker for smaller $\alpha$, the dispersion coefficient is reduced and approaches its high-frequency limit more rapidly. For weak activity, the same trend as in the passive case is observed [$Pe_s=1$; see figure \ref{fig:alpha}(b)]. At higher activity, the swim diffusivity becomes noticeable. For $Pe_s=5$ [see figure \ref{fig:alpha}(c)], $\avg{D^{\mathrm{eff}}}/D^{\mathrm{eff}}_{\mathrm{nf}}$ is larger for smaller $\alpha$ in the intermediate-$\chi$ regime. Regardless of activity, $\avg{D^{\mathrm{eff}}}/D^{\mathrm{eff}}_{\mathrm{nf}}$ approaches the high-frequency limit faster for lower $\alpha$. Because the high-frequency limit exceeds the low-frequency limit in figure \ref{fig:alpha}(c), $\avg{D^{\mathrm{eff}}}/D^{\mathrm{eff}}_{\mathrm{nf}}$ increases more rapidly for smaller $\alpha$. By contrast, in figures \ref{fig:alpha}(a) and \ref{fig:alpha}(b), the high-frequency limits lie below the corresponding low-frequency limits, so decreasing $\alpha$ instead causes $\avg{D^{\mathrm{eff}}}/D^{\mathrm{eff}}_{\mathrm{nf}}$ to decrease more rapidly.

\section{Concluding remarks}
\label{Sec :Conclusions}
In this paper, we employed a GTD theory to study the longitudinal dispersion of ABPs in oscillatory Poiseuille flow. For passive particles, the time-averaged dispersion coefficient decreases monotonically with increasing flow oscillation frequency. As the frequency increases, Taylor dispersion is gradually suppressed due to the increasing oscillations of the flow. The long-time dispersion can be modeled as a random walk, from which a diffusivity is defined as $\ell_\mathrm{eff}^2/\tau_\mathrm{eff}$, where $\ell_\mathrm{eff}$ is the step length and $\tau_\mathrm{eff}$ is the de-correlation time.  In the high-frequency limit, the step length $\ell_\mathrm{eff}$ vanishes as a result of the rapid back-and-forth motion induced by the flow. Therefore, Taylor dispersion vanishes and $\langle D^\mathrm{eff}\rangle \to D_\mathrm{nf}^\mathrm{eff} = D_T$ as $\chi \to \infty$. For active particles, we have shown that the high-frequency behavior is indistinguishable from that of passive particles when the scaled dispersion coefficient $\langle D^\mathrm{eff}\rangle /D_\mathrm{nf}^\mathrm{eff}$ is considered. We note that for active particles $D_\mathrm{nf}^\mathrm{eff} =D_T + D^\mathrm{swim}>D_T$.

We have shown that the effective dispersion coefficient of active particles can exhibit oscillatory behavior as a function of the flow frequency $\chi$. When the external driving frequency (i.e., flow oscillation frequency) matches an intrinsic frequency, resonant diffusion can be observed. This distinct behavior of active particles results from the coupling between self-propulsion and oscillatory fluid advection. Without activity, resonant diffusion does not occur. Likewise, for active particles in a steady Poiseuille flow, no oscillatory dispersion arises due to the absence of a periodic driving force. In oscillatory Poiseuille flow, the oscillation frequency acts as an external control parameter that modulates particle dispersion.  This modulation is particularly versatile for active particles, as flow oscillations can either enhance or suppress dispersion compared to the no-flow case.

In general, resonant or oscillatory dynamics may occur when multiple transport mechanisms are present. 
For example, the rotational dispersion coefficient of axisymmetric Brownian particles in oscillatory shear flows exhibits oscillatory behavior as a function of the flow frequency \citep{leahy2015effect}. In this case, the natural frequency corresponds to the inverse of half a Jeffery orbit period, while the external frequency is the flow oscillation frequency. Resonant diffusion has also been observed in particle systems, including the diffusion of chiral particles in steady Poiseuille flow \citep{khatri2022diffusion}, gravitactic circle
swimmers \citep{chepizhko2022resonant}, and particles in complex time-periodic flow fields with mean flow \citep{Castiglione_1998}.

As is typical in Taylor dispersion theory, we have treated the active particles as point particles, neglecting hydrodynamic interactions with the channel walls. We note that for active particles, their run length $\ell$  is often more relevant than the particle size $a$, particularly when $\ell  \gg a$. When the particle size becomes comparable to the channel width, hydrodynamic interactions can no longer be ignored. Wall-induced hydrodynamic effects may qualitatively alter the effective dispersion. Moreover, active particles generate disturbance flows due to their intrinsic force dipoles \citep{saintillan2014theory}, which can influence both their own motion and that of nearby particles, in contrast to passive particles that only impose no-slip boundary conditions on the surrounding fluid. Incorporating such effects are an important extension of the current model and would be relevant for studying strongly confined swimmers or dense suspensions in microfluidic channels.

In our model, we have assumed that the active particles undergo both translational and rotational Brownian motion. An interesting limit case arises when translational diffusion is absent ($D_T \equiv 0$), such that the active particles are only subject to rotational noise. If $D_T$ is set to zero from the outset, the no-flux boundary conditions [see equations \eqref{eq:BC-1} and \eqref{eq:BC-2}] can no longer be satisfied. To resolve this difficulty, it is necessary to retain $D_T$ in the governing equations and consider a singular perturbation analysis in the limit $D_T \to 0$ \citep{peng2020upstream}. In non-dimensional quantities, one can consider the limit $\gamma \to 0$ (see \S~\ref{Subsec :Non-dimensionalization}). Briefly, diffusive boundary layers will form at both walls, where the diffusive flux balances the swimming flux in order to satisfy the no-flux condition. Alternatively, BD simulations can be used to characterize the dispersion dynamics in the case of zero $D_T$. In BD, translational diffusivity can be set directly to zero without introducing any difficulty. For steady Poiseuille flows, active particles without translational Brownian motion are shown to exhibit giant Taylor dispersion, where $D^\mathrm{eff}/{(U_s^2\tau_R/2)} \sim Pe^4$ in the strong flow limit \citep{peng2020upstream}. From this, we expect that the dispersion of active particles in oscillatory Poiseuille flow without translational noise to exhibit qualitatively different behavior compared with the case of finite $D_T$.

While the GTD theory applies to generic time-periodic flows, we have considered only the case where the driving pressure gradient consists of a single harmonic. An interesting extension would be to include a mean flow, in addition to the oscillatory component that averages to zero. Particle transport due to the interaction between the steady and oscillatory components of the flow may lead to qualitatively different dispersion behavior. In particular, it would be interesting to examine how the oscillatory dispersion behavior is modified.  
    Furthermore, future work should also explore the dynamics of active particles in generic oscillatory flows composed of multiple harmonics, as coupling between different oscillation modes may give rise to qualitatively new dispersion dynamics.
 While our analysis focuses on flows in planar channels, the GTD theory can be generalized to corrugated channels and periodic porous media \citep{peng2024macrotransport,alonso2019transport}. For example, it would be interesting to examine the transport behavior of active particles in peristaltic flow \citep{chakrabarti2020shear}.

    Finally, we note that the present analysis has focused on the long-time, asymptotic dispersion characterized by generalized Taylor dispersion theory. At short times, however, particle transport is governed by pre-asymptotic or transient dispersion, during which the effective longitudinal spreading can differ from its long-time behavior,  and retain memory of the initial conditions and the oscillatory forcing. Transient dispersion has been examined previously for passive and active particles \citep{mukherjee1988dispersion,salles1993taylor,phillips1997initial,bandyopadhyay1999unsteady,vedel2012transient,jiang2021transient,wang2022transient,guan2023pre,zeng2024transient}. An important direction for future work is to consider the transient dispersion of active particles in oscillatory flows.

\section*{Funding}
Numerical simulations were enabled in part by the Digital Research Alliance of Canada. Z. P. was supported by the Discovery Grants program (Grant No. RGPIN-2025-05310) of the Natural Sciences and Engineering Research Council of Canada (NSERC).  Z. P. acknowledges support from the Erskine Fellowship at the University of Canterbury, during which part of this work was completed. 

\section*{Declaration of interests}
 The authors report no conflict of interest.

\appendix

\section{The passive solution} 
\label{app :passive-solution}
The displacement field at $O(1)$ is governed by 
\begin{subequations}
 \begin{equation}
    \frac{\partial b_0^*}{\partial t^*} + \frac{\partial }{\partial y^*}\left(  - \gamma^2 \frac{\partial b_0^*}{\partial y^*}\right) +\frac{\partial }{\partial\phi}\left( \Omega^*_f b_0^* - \frac{\partial b_0^*}{\partial \phi}\right)=\left(U_0^{\mathrm{eff}*} - u^*\right) g_0,
\end{equation}
\begin{equation}
     - \frac{\partial b_0^*}{\partial y^*}=0,\quad\mathrm{at}\quad y^*=\pm 1, 
\end{equation}
\begin{equation}
    \int_{-1}^1 \odiff y^*\intS b^*_0 \dq =0,
\end{equation}
\end{subequations}
which admits a solution of the form $b_0^* = \Re[A_0^\prime(y^*) e^{i\chi t^*} /(2\pi)]$. Here $A_0^\prime$ satisfies 
\begin{subequations}
\begin{equation}
    i\chi A_0^\prime -  \gamma^2\frac{\odiff^2 A_0^\prime}{\odiff y^{*2}} =\overline{u^{\prime *}} - u^{\prime *},
\end{equation}
\begin{equation}
- \frac{\odiff A_0^\prime}{\odiff y^*}=0,\quad\mathrm{at}\quad y^*=\pm 1,  
\end{equation}
\begin{equation}
    \int_{-1}^1 A_0^\prime\odiff y^*=0. 
\end{equation}
\end{subequations}
One can show that the solution is given by 
\begin{equation}
\label{eq:A-0-sol}
     A_0^{\prime}(y^{*}) = \alpha_0 + \alpha_1 \cosh{\left((1+i)\frac{\sqrt{\chi}}{\sqrt{2}\gamma}y^{*}\right) } + \alpha_2 \cosh{((1+i)\kappa y^{*})},
\end{equation}
where 
\begin{subequations}
    \begin{equation}
        \alpha_0 = \frac{Pe(1-i)}{2\kappa^3\chi}\tanh{((1+i)\kappa\,)},
    \end{equation}
    \begin{equation}
        \label{eq: alpha1}
        \alpha_1 = -\frac{\sqrt{2} Pe\, \gamma }{\chi ^{1/2} \kappa  \left(2\kappa^2\gamma^2-\chi\right)}\frac{\tanh \left((1+i)\kappa  \right)}{\sinh(1+i)\frac{\sqrt{\chi}}{\sqrt{2}\gamma}}, 
    \end{equation}
    \begin{equation}
        \label{eq: alpha2}
        \alpha_2= \frac{Pe}{ \kappa ^2 \left(2\kappa^2\gamma^2-\chi\right)}\sech\left((1+i) \kappa\right).
    \end{equation}
\end{subequations}
One interesting limit is $\kappa \to 0$, where the viscous length scale, $\sqrt{2\nu/\omega}$,  is much larger than the channel half-width, $H$. As $\kappa \to 0$, we have 
\begin{subequations}
    \begin{equation}
        \alpha_0 =  \frac{Pe}{\chi\kappa^2} + O(\kappa^2), \quad \alpha_2 = -\frac{Pe}{\chi\kappa^2} + O(\kappa^2),
    \end{equation}
    \begin{equation}
        \alpha_1 = \frac{(1+i) \sqrt{2} Pe }{\chi ^{3/2}} \frac{1}{\sinh\left((1+i) \frac{\sqrt{\chi}}{\sqrt{2}\gamma}\, \right)}+O(\kappa^2),
    \end{equation}
    \begin{equation}
        \cosh{((1+i)\kappa y^{*})} = 1 + O(\kappa^2).
    \end{equation}
\end{subequations}
Therefore, the singular contributions from $\alpha_0$ and $\alpha_2\cosh[(1+i)\kappa y^*]$ in \eqref{eq:A-0-sol} are canceled out while the second term in \eqref{eq:A-0-sol} is regular. Overall $A^{\prime}_0(y^{*})$ is finite as $\kappa \rightarrow 0$, 
\begin{align}
    A^{\prime}_0(y^{*}) &= \frac{Pe}{3\chi^2} 
    \left( -6\gamma^2 + i\left(1 - 3y^{*2}\right)\chi \right) \notag \\
    &\quad + \frac{\sqrt{2}Pe\,\gamma\left(1 + i\right)}{\chi^{3/2}} 
    \frac{\cosh{\left((1+i)\frac{\sqrt{\chi}}{\sqrt{2}\gamma}y^{*}\right) }}
    {\sinh{\left((1+i)\frac{\sqrt{\chi}}{\sqrt{2}\gamma}\right) }} + O(\kappa^2).
\end{align}

Another limit that we examine is $\chi \rightarrow 0$. In this limit, we have
\begin{subequations}
    \begin{equation}
    \alpha_0 = \left( \frac{1}{2} - \frac{i}{2} \right) \frac{Pe}{\kappa^3 \chi} \tanh\left((1 + i)\kappa\right) + O(\chi^3), 
    \end{equation}
    \begin{equation}
        \alpha_1 =  -\left( \frac{1}{2} - \frac{i}{2} \right) \frac{Pe}{\kappa^3 \chi} \tanh\left((1 + i)\kappa\right) + O(\chi), 
    \end{equation}
    \begin{equation}
        \alpha_2 = \frac{Pe}{2\gamma^2\kappa^4}\sech\left(
        (1 + i)\kappa
        \right) + O(\chi),
    \end{equation}
    \begin{equation}
        \cosh{\left((1+i)\frac{\sqrt{\chi}}{\sqrt{2}\gamma}y^{*}\right) }  = 1 + O(\chi).
    \end{equation}
\end{subequations}
The singular contributions from $\alpha_0$ and $\alpha_1\cosh{\left((1+i)\frac{\sqrt{\chi}}{\sqrt{2}\gamma}y^{*}\right) } $ in \eqref{eq:A-0-sol} are canceled out while the third term in \eqref{eq:A-0-sol} is regular. Overall $A^{\prime}_0(y^{*})$ is finite as $\chi \rightarrow 0$,

\begin{align}
    A^{\prime}_0(y^{*}) &= \frac{Pe}{2\gamma^2\kappa^4} 
    \frac{\cosh\left( (1 + i)y^{*}\kappa \right)}
    {\cos\left( (1 + i)\kappa \right)}  \notag \\
    &\quad + \frac{(1 + i)Pe}{12\gamma^2\kappa^5} 
    \left[
    3i + \left( 1 - 3y^{*2} \right) \kappa^2
    \right] \tanh\left( (1 + i)\kappa \right) + O(\chi).
\end{align}

The last limit that we consider is $\left(2\kappa^2\gamma^2-\chi\right) \rightarrow 0$. We define $\epsilon = \left(2\kappa^2\gamma^2-\chi\right)$. This limit $\epsilon \rightarrow 0$ is of particular interest when we look at \eqref{eq: alpha1} and \eqref{eq: alpha2}. We show that in this limit, 
\begin{subequations}
\label{eq: eps}
    \begin{equation}
        \alpha_0 = \left(
        \frac{1}{4} - \frac{i}{4}
        \right)\frac{Pe}{\gamma^2\kappa^5} \tanh\left(
        (1 + i)\kappa
        \right) + O(\epsilon),
    \end{equation}
    \begin{equation}
        \alpha_1 = -\frac{Pe}{\kappa^2\epsilon}\sech\left(
        (1 + i)\kappa
        \right) + O(\epsilon),
    \end{equation}
    \begin{equation}
        \alpha_2 = \frac{Pe}{\kappa^2\epsilon}\sech\left(
        (1 + i)\kappa
        \right) + O(\epsilon^3),
    \end{equation}
    \begin{equation}
        \cosh{\left((1+i)\frac{\sqrt{\chi}}{\sqrt{2}\gamma}y^{*}\right) }  = 
        \cosh{\left((1 + i)\kappa y^{*} \right)} + O(\epsilon).
    \end{equation}
\end{subequations}
The singular contributions from $\alpha_1 \cosh{\left((1+i)\frac{\sqrt{\chi}}{\sqrt{2}\gamma}y^{*}\right) }$ and $\alpha_2 \cosh{((1+i)\kappa y^{*})}$ are canceled out while the first term in \eqref{eq:A-0-sol} is regular. This shows in the limit 
$\epsilon \rightarrow 0$, $A^{\prime}_0(y^{*})$ is finite, and $A^{\prime}_0(y^{*})$ takes the form 
\begin{equation}
    A^{\prime}_0(y^{*}) = \eta_0 + \eta_1\cosh\left(
    (1 + i)y^{*}\kappa
    \right) + \eta_2\sinh\left(
    (1 + i)y^{*}\kappa
    \right) + O(\epsilon),
\end{equation}
where
\begin{subequations}
    \begin{equation}
        \eta_0 = \frac{(1 - i)Pe}{4\gamma^2\kappa^5}\tanh\left(
    (1 + i)\kappa
    \right) - \frac{Pe}{4\gamma^2\kappa^4}\frac{1}{\sinh\left(
    (1 + i)\kappa
    \right)},
    \end{equation}
    \begin{equation}
        \eta_1 = 1 + (1 + i)\kappa \coth\left(
    (1 + i)\kappa
    \right),
    \end{equation}
    \begin{equation}
        \eta_2 = 
    - y^{*}(1 + i)\kappa\tanh\left(
    (1 + i)\kappa
    \right).
    \end{equation}
\end{subequations}

\section{The high-frequency limit}
\label{app :high-frequency}
Here we analyze the governing equations \eqref{eq:g-non-dimensional} and \eqref{eq:b-non-dimensional} in the high-frequency limit characterized by  $\chi \gg 1$ while keeping the ratio $\chi/\kappa^2 =\alpha= 2\nu\tau_R/H^2$ constant. That is, $\alpha = O(1)$ as $\chi \to \infty$.  We also assume that all other non-dimensional parameters are $O(1)$ as $\chi \to \infty$. To facilitate our analysis, we write the long-time solution to $g$ as a Fourier series, 
\begin{equation}
    g(y^*, \phi, t^*) = \sum_{n=-\infty}^{+\infty}e^{i n \chi t^*} g_n(y^*, \phi). 
\end{equation}
Inserting this expansion into equation \eqref{eq:g-non-dimensional}, we obtain 
\begin{align}
    &i n \chi g_n+ Pe_s \sin\phi\, \frac{\partial g_n}{\partial y^*} - \gamma^2 \frac{\partial^2 g_n}{\partial {y^*}^2} \nonumber \\
    & + \frac{\partial }{\partial \phi}\left[ \text{Re}\left(\Omega^{\prime *}\right)\frac{1}{2}\left(g_{n-1}+g_{n+1}\right) - \text{Im}\left(\Omega^{\prime *}\right)\frac{1}{2i}\left(g_{n-1}-g_{n+1}\right)\right] - \frac{\partial^2 g_n}{\partial \phi^2}=0.
\end{align}
The conservation condition becomes
\begin{equation}
    \frac{1}{2}\int_{-1}^1 \dy^*\int_0^{2\pi}\dphi\sum_{n=-\infty}^{+\infty} e^{in\chi t^*}g_n(y^*, \phi)=1.
\end{equation}
Making use of orthogonality, we have 
\begin{equation}
    \frac{1}{2}\int_{-1}^1 \dy^*\int_0^{2\pi}\dphi \, g_0(y^*, \phi)=1, \quad \text{and} \quad \int_{-1}^1 \dy^*\int_0^{2\pi} \dphi \,g_n(y^*, \phi)=0, n\neq 0.
\end{equation}
In the high-frequency limit, one can show that $g_0 = O(1)$ and $g_1 = o(1)$. At leading order, we have 
\begin{equation}
Pe_s \sin\phi\, \frac{\partial g_0}{\partial y^*} - \gamma^2 \frac{\partial^2 g_0}{\partial {y^*}^2}  - \frac{\partial^2 g_0}{\partial \phi^2}=0.
\end{equation}
The no-flux condition is given by 
\begin{equation}
    Pe_s \sin\phi\, g_0 - \gamma^2 \frac{\partial g_0}{\partial {y^*}}=0, \quad y^*=\pm 1. 
\end{equation}
This shows that, at high frequencies, the governing equation for $g$, to leading order, reduces to that of ABPs in a channel without flow.
Similarly, one can show that the displacement field also satisfies the equation without flow. As a result, the effective dispersion coefficient approaches the no-flow result in the high-frequency limit.

\section{Brownian dynamics} 
\label{app :BD}
The discretized Langevin equations are given by 
\begin{subequations}
    \begin{equation}
        x_{n+1} =  x_n + u_f\left(y_n,t_n\right)\Delta t + U_s\cos\left(\phi_n\right)\Delta t + \Delta x^B,
    \end{equation}
    \begin{equation}
        y_{n+1} = y_n + U_s\sin\left(\phi_n\right)\Delta t + \Delta y^B,
    \end{equation}
    \begin{equation}
        \phi_{n+1} = \phi_n + \Omega_f(y_n, \phi_n, t_n) \Delta t + \Delta \phi^B,
    \end{equation}
\end{subequations}
where $\Delta t$ is the time step.  The Brownian displacements  $\Delta x^B$,  $\Delta y^B$, and $\Delta \phi^B$ are sampled from independent white noise processes. The translational Brownian displacement has a variance of $2D_T\Delta t$, and the rotary Brownian displacement has a variance of $2\Delta t/\tau_R$. A potential-free algorithm is used to implement the no-flux condition \citep{heyes1993brownian}. For all simulations, a sufficiently small time step is used to resolve all the physical timescales in the system. The total simulation time is long enough to ensure convergence to the long-time behavior.  A timestep in the range $10^{-6} \tau_R $--$10^{-5} \tau_R$ is used.  To ensure good statistics, all simulations are performed with $200,000$ particles.

\begin{figure} 
    \centering
    \includegraphics[width=5in]{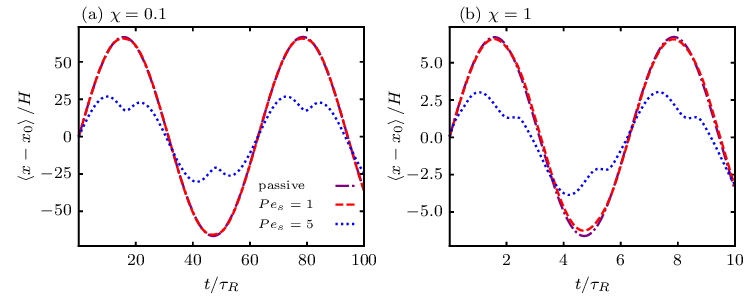}
    \caption{ 
        Plots of the  mean displacement in the $x$ direction, $\langle x - x_0 \rangle/H$ as a function of $t/\tau_R$ for different values of $Pe_s$ for (a) $\chi = 0.1$ and (b) $\chi = 1$.  All results are from BD simulations with $\alpha = 100$, $\gamma^2 = 0.1$, and $Pe=10$. 
      }
    \label{fig:meanxdisplacement} 
\end{figure}

\begin{figure} 
    \centering
    \includegraphics[width=5in]{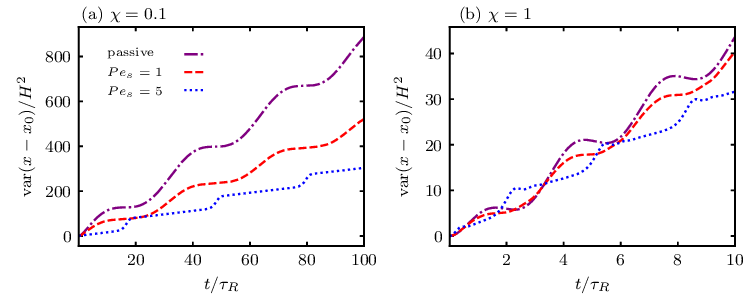}
    \caption{ 
        Plots of the variance in the $x$ direction, $\mathrm{var}(x - x_0)/H^2$, as a function of $t/\tau_R$ for different values of $Pe_s$ for (a) $\chi = 0.1$, and (b) $\chi = 1$. All results are from BD simulations with $\alpha = 100$, $\gamma^2 = 0.1$, and $Pe=10$.
      }
    \label{fig:varx} 
\end{figure}

    In figure \ref{fig:meanxdisplacement}, we show the transient behavior of the mean displacement $\langle x - x_0 \rangle/H$ as a function of $t/\tau_R$, for different values of $\chi$ and for different values of $Pe_s$. Here $x_0$ is the initial $x$ positions of particles. At long times, the mean displacement follows the oscillatory behavior of the flow, as it oscillates around zero. In figure \ref{fig:varx}, we show the transient behavior of  the variance $\mathrm{var}(x - x_0)/H^2$ as a function of $t/\tau_R$, for different values of $\chi$ and for different values of $Pe_s$. The variance grows with time while exhibiting oscillations that reflect the time-periodic nature of the flow.

\section{Numerical simulation}
\label{app :Numerical}
The governing equations \eqref{eq:g-non-dimensional} and \eqref{eq:b-non-dimensional} are solved numerically using Dedalus \citep{Burns2020}. The physical space ($y^*$) is discretized on a Chebyshev grid with $128$ nodes, and the orientational space is represented in Fourier space with $128$ nodes. For time integration, we employ a second-order Crank-Nicolson-Adams-Bashforth scheme.  A typical timestep  of $10^{-6 }$ is used. In the high-frequency regime, the timestep is further reduced to $10^{-7}$.

\bibliographystyle{jfm}
\bibliography{refs}

\end{document}